\documentclass[sigconf]{acmart}
\usepackage{array} 
\usepackage{float} 
\usepackage{soul} 
\usepackage{longtable} 
\usepackage{colortbl} 
\usepackage{stfloats} 

\AtBeginDocument{%
  \providecommand\BibTeX{{%
    \normalfont B\kern-0.5em{\scshape i\kern-0.25em b}\kern-0.8em\TeX}}}

\copyrightyear{2023}
\acmYear{2023}
\setcopyright{rightsretained}
\acmConference[DIS '23]{Designing Interactive Systems Conference}{July 10--14, 2023}{Pittsburgh, PA, USA}
\acmBooktitle{Designing Interactive Systems Conference (DIS '23), July 10--14, 2023, Pittsburgh, PA, USA}
\acmDOI{10.1145/3563657.3596059}
\acmISBN{978-1-4503-9893-0/23/07}

\begin{document}

\title[Rewriting the Script: Adapting Text Instructions for Voice Instruction]{Rewriting the Script: \\ Adapting Text Instructions for Voice Interaction}

\settopmatter{authorsperrow=4}

\author{Alyssa Hwang}
\email{ahwang16@seas.upenn.edu}
\affiliation{%
  \institution{University of Pennsylvania}
  \streetaddress{3300 Walnut St.}
  \city{Philadelphia}
  \state{PA}
  \country{USA}
  \postcode{19104}
}

\author{Natasha Oza}
\email{noza@sas.upenn.edu}
\affiliation{%
  \institution{University of Pennsylvania}
  \streetaddress{3300 Walnut St.}
  \city{Philadelphia}
  \state{PA}
  \country{USA}
  \postcode{19104}
}

\author{Chris Callison-Burch}
\email{ccb@seas.upenn.edu}
\affiliation{%
  \institution{University of Pennsylvania}
  \streetaddress{3300 Walnut St.}
  \city{Philadelphia}
  \state{PA}
  \country{USA}
  \postcode{19104}
}

\author{Andrew Head}
\email{head@seas.upenn.edu}
\affiliation{%
  \institution{University of Pennsylvania}
  \streetaddress{3300 Walnut St.}
  \city{Philadelphia}
  \state{PA}
  \country{USA}
  \postcode{19104}
}

\renewcommand{\shortauthors}{Hwang et al.}

\begin{abstract}
Voice assistants have sharply risen in popularity in recent years, but their use has been limited mostly to simple applications like music, hands-free search, or control of internet-of-things devices. What would it take for voice assistants to guide people through more complex tasks? In our work, we study the limitations of the dominant approach voice assistants take to complex task guidance: reading aloud written instructions. Using recipes as an example, we observe twelve participants cook at home with a state-of-the-art voice assistant. We learn that the current approach leads to nine challenges, including obscuring the bigger picture, overwhelming users with too much information, and failing to communicate affordances. Instructions delivered by a voice assistant are especially difficult because they cannot be skimmed as easily as written instructions. Alexa in particular did not surface crucial details to the user or answer questions well. We draw on our observations to propose eight ways in which voice assistants can ``rewrite the script''---summarizing, signposting, splitting, elaborating, volunteering, reordering, redistributing, and visualizing---to transform written sources into forms that are readily communicated through spoken conversation. We conclude with a vision of how modern advancements in natural language processing can be leveraged for intelligent agents to guide users effectively through complex tasks.
\end{abstract}



\newcommand{\todo}[1]{\out{{\small\textcolor{purple}{\bf [*** TODO: #1]}}}}
\newcommand\alyssa[1]{\textcolor{blue}{[AH: #1]}}
\newcommand\chris[1]{\textcolor{brown}{[CCB: #1]}}

\definecolor{andrewpurple}{HTML}{A53DFF}
\definecolor{andreworange}{HTML}{E07400}
\definecolor{darkgreen}{HTML}{009B55}

\newcommand\andrew[1]{\textcolor{andrewpurple}{#1}}
\newcommand\important[1]{\textcolor{darkgreen}{#1}}
\newcommand\unimportant[1]{\textcolor{gray}{\sout{#1}}}
\newcommand\move[1]{\textcolor{andreworange}{#1}}

\definecolor{niceblue}{HTML}{8295ff}
\def\bigbox{\color{niceblue}\rule[.25ex]{1ex}{1ex} \hskip .1ex}
\def\smallbox{\hskip .25ex \color{niceblue}\rule[.5ex]{.5ex}{.5ex} \hskip .25ex \hskip .1ex}
\def\boxes#1#2{
\hskip .1ex 
\newcount\boxnum
\boxnum=0
\loop
\ifnum \boxnum<#1 \bigbox \else \smallbox \fi
\advance \boxnum by 1
\ifnum \boxnum<#2
\repeat
}

\def\shortspace{\kern 0.1em}

\def\alexa#1{{\small\texttt{#1}}}

\maketitle

\section{Introduction}
\label{sec:introduction}
Voice assistants have become very popular in recent years, turning what was once science fiction into an everyday reality. As voice assistants have matured, they have started to be used in cases of growing complexity and nuance, like helping researchers perform procedures in the lab~\citep{cambre_vitro_2019} and individuals with cognitive disabilities perform self-care activities at home~\citep{carroll_robin_2017}. In many cases, voice assistants have the potential to make information more accessible, but they remain used primarily for simple tasks: \citet{ammari_music_2019} found that the predominant uses of Amazon Alexa and Google Home are centered around features like music, hands-free search, and control of internet-of-things devices. Furthermore, voice assistants have been critiqued for their limited ability to carry on conversations with the same flexibility that we expect from conversations with people~\citep{porcheron_voice_2018}. It seems like we have not yet discovered the recipe for supporting longer-form interactions with a voice assistant.

In this paper, we explore how voice assistants can effectively support one kind of longer-form interaction: guiding users through complex tasks. From lab procedures to recipes, people frequently work on tasks that could benefit from hands-free, eyes-free guidance. Most voice assistants currently guide users through complex tasks by reading aloud written instructions one step at a time with few modifications. This approach benefits from the widespread availability of written instructions, but instructions written for the page do not always transfer well directly to audio. Adapting instructions into audio-friendly forms is important for all intelligent agents that communicate through spoken conversation, even those with smart displays or other visual output modalities.

To learn more about the limitations of this approach to complex task guidance, we conducted an observational study of 12 individuals using recipes as an example. These participants, representing a range of cooking skill and enthusiasm for experimenting with voice assistants, worked with Amazon Alexa to prepare recipes of varying complexity. The results of this study are an account of 9 challenges users face when following a recipe with Alexa: missing the big picture, information overload, fragmentation, time insensitivity, missing details, discarded context, failure to listen, uncommunicated affordances, and limitations of audio (see Table~\ref{tab:challenges}). The core source of these challenges seemed to be the inability to navigate through a recipe with Alexa as easily as they would have been able to skim through a written one. Since Alexa dictated steps one at a time, some crucial details further along in the recipe were nearly impossible to find or anticipate. Alexa did not surface these details to the participants or answer questions well, either. It even failed to respond to participants' attempts to interact with it on many occasions, whether or not they used the wake word.

We suggest addressing these challenges by designing voice assistants to ``rewrite the script'': adapt written instructions into a form that is easier to follow in hands- and eyes-free settings. In our discussion, we outline a set of capabilities involved in rewriting the script: summarize, signpost, split, elaborate, volunteer, reorder, redistribute, and visualize (see Table~\ref{tab:hci_nlp_refs}). These capabilities revolve around ways that voice assistants can rearrange information to communicate more effectively with their users. Furthermore, many of these capabilities are already possible with the current state of natural language processing research, especially in task-oriented dialogue, event reasoning, and commonsense reasoning. Given the complementary advances in natural language processing and human-computer interaction, science fiction continues to become a reality at a fast pace. We conclude with a vision of what it might mean for this kind of voice assistant to become part of that reality.

\section{Background and Related Work}
\label{sec:bg_and_rw}
In this section, we review research that offers insight on designing voice assistants for complex task guidance, including voice assistant design, instruction design, and task interfaces.

\subsection{Designing Voice Assistants}
The human-AI interaction community has developed several sets of guidelines for designing good voice assistants. In their landmark paper, \citet{amershi_guidelines_2019} propose eighteen heuristics for designing AI-infused systems, including that they should indicate what they can do and how well they can do it. Notably, voice assistants are known for not communicating their affordances well~\citep{sciuto_hey_2018}. Although voice assistants are relatively new, early work in the 1990s warned that voice interfaces should be ``designed from scratch, rather than directly translated from their graphical counterparts'' \citep{yankelovich_speechacts_1994, yankelovich_designing_1995}. \citet{sherwani_voicepedia_2007}'s later work on VoicePedia echoes this warning: this voice user interface (VUI) mimicked Wikipedia's graphical user interface (GUI) as closely as possible and was rejected in user studies. Since then, contemporary researchers have tackled voice interface design in a new way: transforming existing resources specifically for audio rather than treating VUIs as spoken GUIs~\citep{murad_design_2018, murad_finding_2021}. We follow~\citet{murad_i_2019}'s lead in establishing usability principles for voice interaction from the ground up.

Additional guidance for designing voice assistants focuses on their abilities as conversational agents. \citet{langevin_heuristic_2021}'s heuristics for conversational agents emphasize the need to guide users through the available affordances without overwhelming them. \citet{clark_what_2019} dive even deeper into the meaning of a good conversation, suggesting that conversational agents concentrate on functional rather than social goals. \citet{volkel_eliciting_2021} similarly find that envisioned conversational agents were just social enough to support highly interactive, multi-turn conversations while helping users with a task without becoming a ``friend.'' The design space for conversational agents is large and complex since users can communicate a wide range of intents in many ways~\citep{chang_how_2019, zhao_rewind_2022} and contexts, like while driving~\citep{large_lessons_2019}. Our work focuses on cooking with a voice assistant, reiterating the need to make affordances clearer and support long-form conversations for complex task guidance.

\subsection{The Design of Instructions}
Previous findings in instruction design and cognitive science can help inspire the design of voice interfaces for complex task guidance. One classic result in cognitive science famously suggests that working memory is limited to ``seven, plus or minus two,'' items \citep{miller_magical_1956}. This implies that instructions should limit the amount of information that a user has to keep track of at a time. Following instructions delivered over audio poses unique challenges because verbal instructions are processed by the phonological loop in the brain~\citep{dunham_psychology_2020}. While the phonological loop is faster and more flexible than the structures for visual processing, the information in it decays more rapidly~\citep{dunham_psychology_2020}. Voice assistants need to be particularly strategic about the level of detail provided in any one instruction to respect the limits of our neurobiology.

Prior work suggests some techniques to offer instructions that are mindful of these limits. Simply replacing written text with spoken text---the primary approach to complex task guidance through voice assistants---is not necessarily the right approach since it has led to detrimental effects in some studies~\citep{tabbers_multimedia_2004}. Rather than reciting text verbatim, one approach is to present concrete, well segmented instructions to help users perform unfamiliar tasks~\citep{jannin_atomized_2019}. Regardless of delivery format, concrete procedures have been shown to improve immediate performance while abstract procedures help with learning and transfer~\cite{eiriksdottir_procedural_2011}. Instruction formats can also embrace minimalism, an approach introduced by~\citet{carroll_nurnberg_1990} that focuses on learning skills as needed rather than all ahead of time. In our study, we focus on how written text may need to be transformed for audio-first interfaces given these insights on cognitive processing, concreteness, and minimalism.

\subsection{Intelligent Cooking Support}
Our paper focuses on recipes as one type of instruction that voice assistants may help users follow. The human-computer interaction  community has broadly explored the design of interfaces to support cooking, many of which intersect with the goals of our work. \citet{chang_recipescape_2018}'s RecipeScape, for instance, helps users interactively browse a collection of recipes. Other tools help people follow along with recipes, such as \citet{kosch_digital_2019}'s digital cooking coach, which provides \textit{in situ} auditory and visual feedback on a cook's tasks.
In a more immersive scenario, \citet{sato_mimicook_2014}'s~MimiCook and \citet{chen_smart_2010}'s ``smart kitchen'' embed step-by-step instructions and nutritional information into kitchen counters and screens. Some interfaces allow users to navigate through video recipes with their voices, which requires voice assistants to understand a range of intents~\citep{zhao_rewind_2022}. Our work explores how users navigate through audio recipes as a case study on voice interaction for any complex task.

\subsection{Complex Task Support}
Along with cooking, the human-computer interaction community has envisioned many ways to help people accomplish a wide variety of tasks by augmenting their workspaces~\citep{schoop_drill_2016, nouri_supporting_2019} and devices~\citep{winkler_alexa_2019}. Conversational agents that provide instructional support---like Vitro, a voice assistant that guides researchers through cell culturing procedures~\citep{cambre_vitro_2019}---are particularly relevant to our work. Iris, on the other hand, is a text-based conversational agent that chains together simple commands to perform complex data science tasks~\citep{fast_iris_2018}. Prior research has also indicated the nuance involved in helping users navigate sets of instructions with a voice interface. \citet{abdolrahmani_towards_2021} propose that voice assistants in complex environments like an airport provide support through short transactions. Other work has suggested that interfaces should support multiple kinds of pauses and jumps \cite{chang_how_2019}, handle implicit conversation cues \cite{vtyurina_exploring_2018}, and support jumps according to both conventional navigation instructions and content-based anchors \citep{zhao_rewind_2022}. Our paper contributes a detailed exploration of the challenges involved in following audio-first task guidance and suggestions to overcome them.

\section{Methods}
\label{sec:methods}
\begin{figure}[]
    \centering
    \includegraphics[width=\columnwidth]{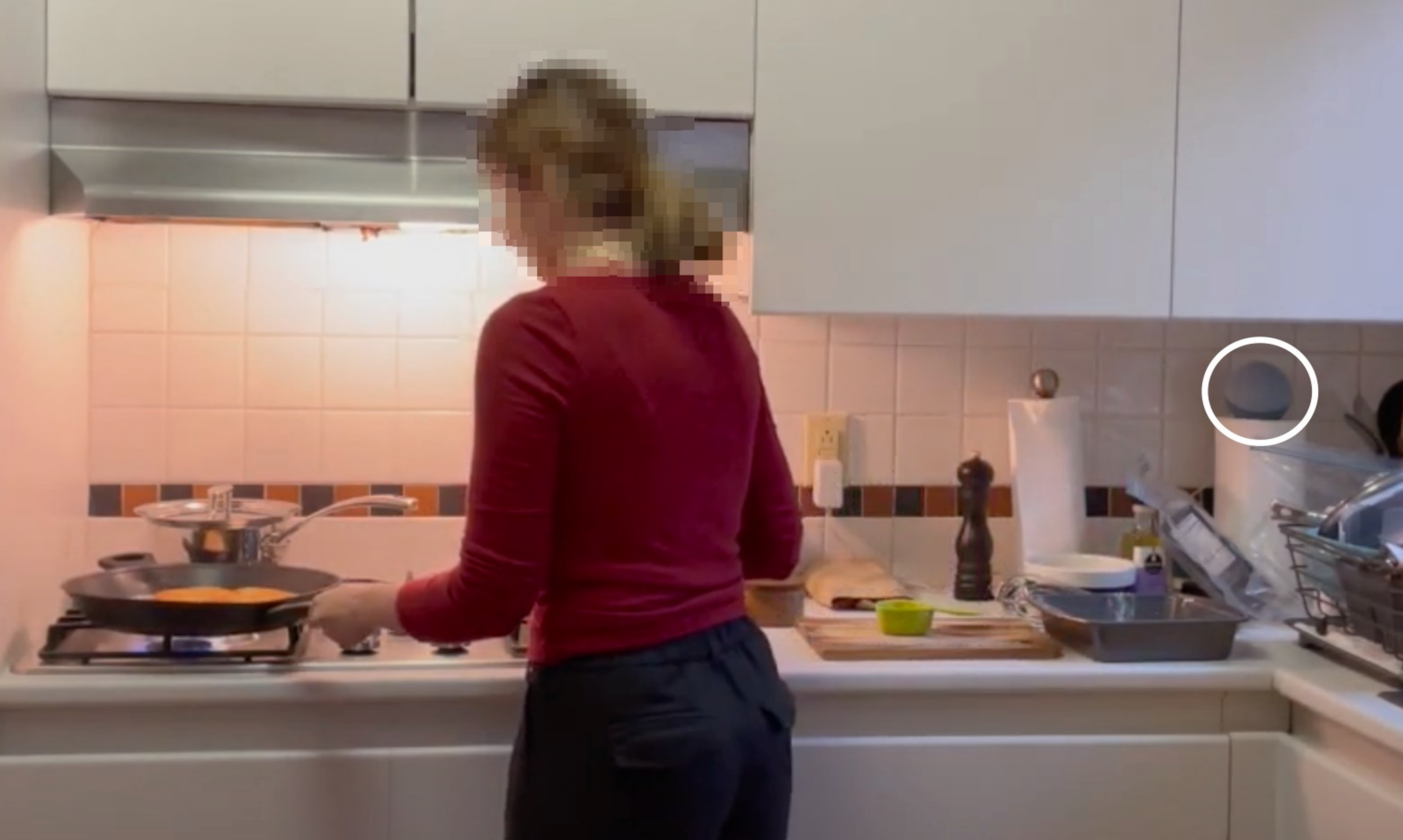}
    \Description[Cooking with Alexa]{One of the cooks preparing a dish with Alexa in the corner of the kitchen.}
    \caption{Study setting. \textmd{Participants followed recipes of their choice with the help of Amazon Alexa (Echo Dot, circled on the right). Participants were observed at home and encouraged to cook however felt natural as we observed, only occasionally asking clarifying questions. We filmed the session with a camera on a tripod out of the way of the kitchen.}}
    \label{fig:cooking_setup}
\end{figure}

\begin{table}[]
\centering
\renewcommand{\arraystretch}{1.2}
\begin{tabular}{l>{\raggedright}p{4cm}>{\centering}p{1.4cm}>{\centering}p{1.2cm}}
\toprule
ID  & Selected Recipe                                                        & Self-Rated Skill            & Prior Use \tabularnewline 

\midrule

C1  & Steaks with Blue Cheese Butter                                & \boxes{4}{5}             & daily                     \tabularnewline
C2  & Eggless Red Velvet Cake                                       & \boxes{4}{5}             & weekly                    \tabularnewline
C3  & Sesame Pork Milanese                                          & \boxes{2}{5}             & \textless{}~monthly        \tabularnewline
C4  & Honey Garlic Chicken Wings                                    & \boxes{3}{5}             & \textless{}~monthly        \tabularnewline
C5  & Teriyaki Salmon                                               & \boxes{1}{5}             & monthly                   \tabularnewline
C6  & Seafood Marinara                                              & \boxes{2}{5}             & never                     \tabularnewline
C7  & Honey Soy-Glazed Salmon                                       & \boxes{4}{5}             & never                     \tabularnewline
C8  & Sausage and Veggie Quiche                                     & \boxes{2}{5}             & daily                     \tabularnewline
C9  & Egg Biryani                                                   & \boxes{5}{5}             & weekly                    \tabularnewline
C10 & Herb-Roasted Salmon with Tomato-Avocado Salsa                     & \boxes{4}{5}             & weekly                     \tabularnewline
C11 & Lebanese Chicken Fatteh                                       & \boxes{2}{5}             & never                    \tabularnewline
C12 & Ground Beef Bulgogi                                           & \boxes{3}{5}             & weekly                    \tabularnewline \bottomrule
\end{tabular}
\Description[Participant information]{The anonymized ID, recipe, self-rated cooking skill, and self-reported frequency of voice assistant use for each cook.}
\caption{Participants. \textmd{Participants were mostly graduate students and chose a wide variety of recipes to prepare. They represented a range of cooking skill (``Self-Rated Skill'' on a 5-point Likert scale) and frequency of voice assistant usage (``Prior Use'').}}
\label{tab:participants}
\end{table}

We designed an observational study to understand how voice assistants can effectively guide people through complex tasks, using recipes as an example. We recruited participants to choose and prepare recipes at home while being guided by a voice assistant (see Figure~\ref{fig:cooking_setup}). We aimed to answer the following research questions:

\begin{quote}
    \textbf{RQ1}: What challenges do users face when following instructions to perform complex tasks given by a contemporary, state-of-the-art voice assistant? \\ 
    \textbf{RQ2}: What can be done to address these challenges in future voice assistants?
\end{quote}

Our goal was to clearly document the challenges in a way that led to concrete suggestions for solutions. To do this, we opted to perform an observational study with deep contextual elements. Even though our study is not a contextual inquiry according to the precise methodology described by \citet[Chapter 3]{beyer_contextual_1997}---participants were not using their own Alexa and they would not have performed the task without our intervention---we made heavy use of contextual elements in the design of the study: we observed participants in natural work settings (their homes) working on tasks they cared about (recipes of their own choice), with continual, incremental interpretation of observations during and after the task. Our hope was that this contextual approach would lead to deep, validated, actionable design inspiration while being possible to arrange in a way that a full contextual inquiry would not be.

\subsection{Technology Probe}

Participants in our study interacted with Amazon Alexa to prepare their recipes. We chose to study Alexa because it was, to our knowledge, the state of the art in hands-free, eyes-free voice interaction. Furthermore, one of the authors of this paper had prior experience working with Alexa for the Alexa Prize Taskbot Challenge, which made us aware of its capabilities for similar tasks \citep{panagopoulou_quakerbot_2022}. We also chose an audio-only device to focus on the design of spoken communication, which voice assistants of all kinds need to support.

We originally used a Wizard-of-Oz approach to represent an idealized version of a voice assistant, but we converged on using Alexa instead because of the challenges associated with developing a realistic, idealized voice assistant for study settings. Existing tools for changing a human's voice to sound more robotic were inappropriate for our goals because most real-time voice changers were designed for humor. Attempting to type responses fast enough to use text-to-speech technology introduced an unnatural 5-to-10--second delay. Our own tests with Alexa revealed that it already provides sophisticated support for complex task guidance, including quickly answering questions with external information, that we felt we could not rival with a WOZ'd prototype. We therefore decided to explore the challenges associated with modern devices and suggest areas for improvement, as revealed by our observational study.

\begin{figure*}[]
    \centering
    \includegraphics[width=\textwidth]{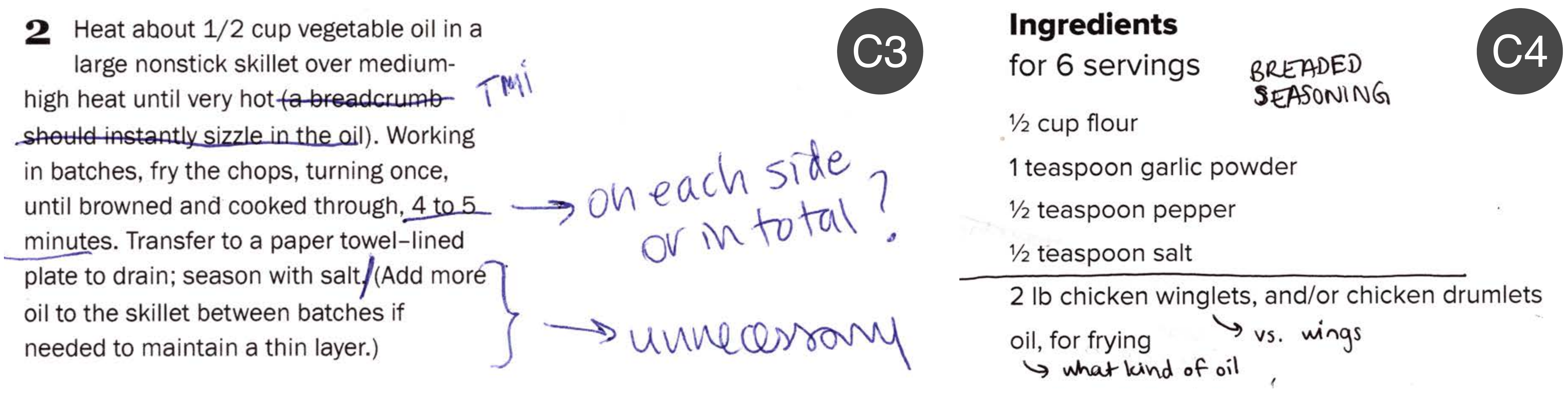}
    \Description[Annotated recipes]{Two examples of written recipes annotated with user suggestions, like removing extraneous information, clarifying details, and splitting steps or ingredients into chunks.}
    \caption{Instructions annotated with user suggestions. \textmd{At the end of each cooking session, cooks were asked to mark up printed copies of the original recipes that they had just prepared with Alexa. Cooks indicated content they wished Alexa had changed for a better audio script, including skipping extra information (see the strikethrough and ``TMI'' for ``too much information,'' C3), providing more details (``on each side or total?'' C3), splitting long steps into multiple shorter substeps (see the ``\big/'' mark, C3), and grouping ingredients into categories (see the ingredients above the line annotated with ``BREADED SEASONING,'' C4).}}
    \label{fig:recipe_excerpt}
\end{figure*}

\subsection{Participants}

Participants were recruited from an institution-wide graduate student email newsletter at the University of Pennsylvania. We chose to scope recruiting to within our university community because we would be observing participants at home. This ensured that participants would be within traveling distance and have some personal connection to the research team, which we believed would make the session more comfortable for participant and researcher alike.

Readers of the newsletter were asked to complete a preliminary questionnaire to indicate their interest, background related to cooking, and prior use of voice assistants. We sampled participants according to two criteria: (1) whether they were available during daytime hours, which we anticipated would make the at-home observation more comfortable, and (2) whether they helped us achieve a wide coverage of cooking and technical experience.

The selected sample of participants varied a great deal in cooking skill and familiarity with voice assistants (see Table~\ref{tab:participants}). Since participants would be completing a task while interacting with a sophisticated piece of technology, we selected for diversity in both areas to learn about a fuller range of experiences. On a 5-point Likert scale (where 5 indicated a great amount of skill), five participants reported their cooking skill at a level of 2 or below, five participants reported 4 or above, and two participants reported exactly 3. Participants also used voice assistants with varying levels of frequency, with two using them daily, four weekly, one monthly, and five less than monthly. Some were excited to experiment with voice assistants, with seven rating their excitement at 4 or 5 out of 5 on a Likert scale; four participants were less excited at a 2 or 3.

As a result of the recruiting method, most cooks were graduate students (67\% Master's, 25\% Ph.D.), with the exception of C5, who was a software engineer.
Eight of the twelve participants answered questionnaire items about their demographic information.\footnote{The question for race was adapted from the 2022 Computing Research Association Annual Survey. The question for ability status was adapted from the Voluntary Self-Identification of Disability provided by the United States Department of Labor.} Of these eight, 63\% self-identified as female and the rest (37\%) as male. Ages ranged between 22 and 30 years old, with a median age of 23.5 years old. 37.5\% reported their race as Caucasian/European/White, 37.5\% East Asian, 12.5\% South Asian, and 12.5\% Southeast Asian. Except for one cook who described herself as ``intermediate,'' all respondents described themselves as ``proficient'' in English. 

\subsection{Procedure}
Once selected, a participant was asked to complete a few steps to prepare for their cooking session. First, they downloaded the Amazon Alexa app and used it to search for the recipe they wanted to cook. They were required to use the Alexa app because it was the only way to ensure that the recipe they chose would be supported by the device. The participant then shopped for the ingredients before the research team arrived for their session. Participants were inevitably able to see the recipe ahead of time to purchase ingredients, so we asked them to minimize the amount of the recipe they read in advance to reduce the likelihood that they would come to the study with significant prior knowledge.

We met the participant at home at the scheduled time to observe them as they cooked. We briefed them on the study procedures and asked them for their consent to participate. We then set up our equipment. In most cases, we used a fourth-generation Amazon Alexa Echo Dot, which was the newest screen-less Alexa device. Occasionally, the Echo failed to connect to the internet, so we used the Amazon Alexa iOS app on an iPhone 12. Lastly, we set up a camera to record the cooking session and debrief interview.

We started the session with a brief overview of how to use Alexa: navigating to the next step, backtracking to the previous step, and jumping to a specific step. We encouraged them to ask Alexa questions and interact with it however felt natural. The participant then prepared their dish with Alexa. As the participant cooked, the research team observed and asked occasional clarification questions to understand critical incidents. We encouraged everyone to think aloud if they felt comfortable, but most seemed to think aloud only a handful of times during each session.

The participant was asked to complete two remaining activities after finishing their recipe. First, we conducted a semi-structured interview to learn more about their experience during the session, including what was easy and difficult about working with Alexa and their willingness to follow a recipe with a voice assistant again on a 5-point Likert scale. Second, the participant was asked to annotate a printed copy of their recipe, indicating changes to Alexa's audio script that could have improved their experience (see Figure~\ref{fig:recipe_excerpt}). We reviewed the annotated recipe and their thought process with them. Finally, we debriefed the participant and concluded the session. Participants were compensated with a gift card amounting to the cost of ingredients and an additional \$100 USD.

\subsection{Analysis}
\label{sec:analysis}

Our study yielded four kinds of data: audio and video recordings, researcher notes, questionnaire data, and annotated paper recipes. We used \hyperlink{rev.com}{\url{Rev.com}} to transcribe audio recordings.
Notes, transcripts, and annotated recipes were analyzed with a thematic analysis approach~\cite[Chapter 5]{blandford_qualitative_2016}. One author developed a set of codes during an open coding pass, reviewing all of the data. Another author reviewed the codes and all accompanying excerpts. Then, both authors revised the set of codes into a final schema. Codes were grouped into categories roughly corresponding to the 9 challenges in Section~\ref{sec:results}. The former author applied this schema to the data in an axial coding pass, which was validated by the latter author. 

To analyze counts of events (such as the number of navigation requests), transcripts were analyzed once more. The author who had originally defined the set of codes created a code book of event types, along with examples and brief written descriptions of each one. Two authors then applied this code book to every conversational turn in two transcripts. The boundaries of the conversational turns were determined by the transcribers at Rev. After showing high agreement on all codes, one author applied the code book to the remaining transcripts (see Appendix Table~\ref{tab:validation}).

\section{Results}
\label{sec:results}
\begin{figure*}[]
    \centering
    \includegraphics[width=\textwidth]{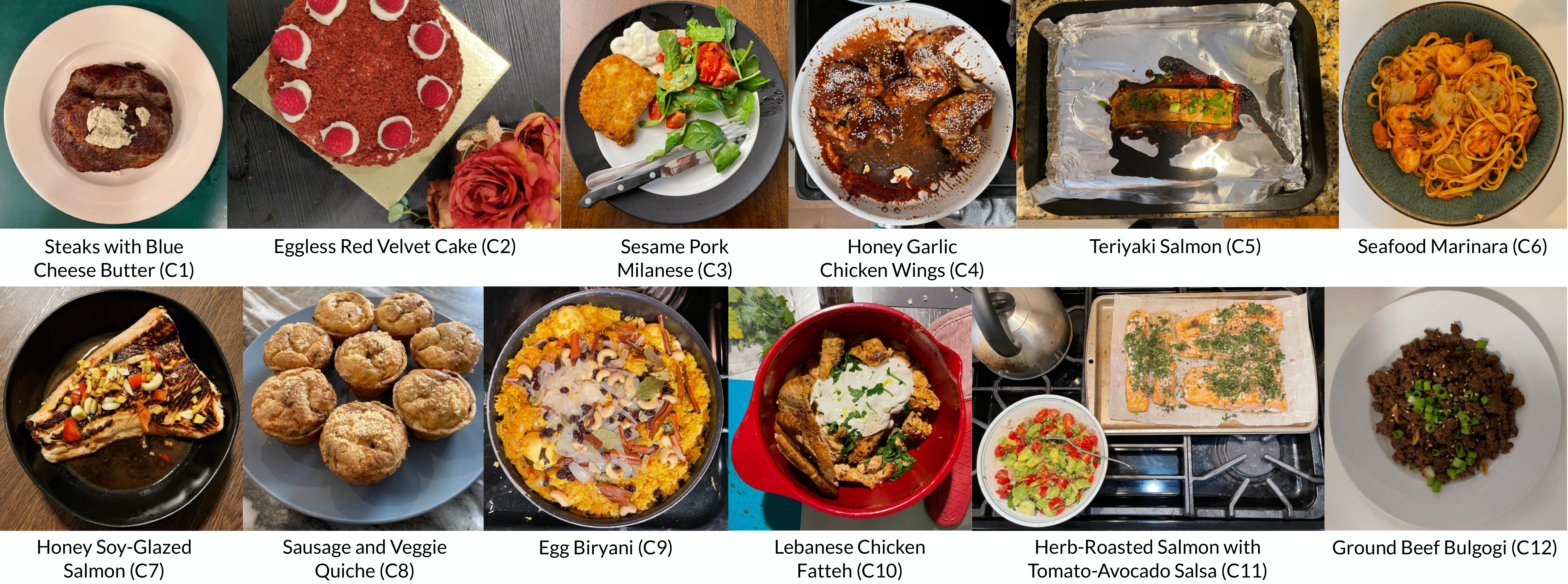}
    \Description[Food photos]{Photos of each cook's finished dishes: steaks with blue cheese butter, eggless red velvet cake, sesame pork Milanese, honey garlic chicken wings, teriyaki salmon, seafood marinara, honey soy-glazed salmon, sausage and veggie quiche, egg biryani, Lebanese chicken fatteh, herb-roasted salmon with tomato-avocado salsa, and ground beef bulgogi.}
    \caption{Completed dishes.
    \textmd{Cooks prepared a variety of dishes of their choice following the guidance of a voice assistant. These dishes varied in complexity: some required interaction with the voice assistant for many steps (i.e., C2's eggless red velvet cake), while others involved just a few (i.e., C12's ground beef bulgogi).}
    }
    \label{fig:food}
\end{figure*}

\begin{table*}[]
\renewcommand{\arraystretch}{1.5}
\begin{tabular}[t]{l>{\raggedright}p{3cm}>{\raggedright}p{3.6cm}p{9.2cm}}
\toprule
\# &  Challenge               & Description                                                                                & Representative Observation                                                                                                                                                                                                                                                                                                                                                                                        \\ \midrule
1 & Missing the Big Picture &
Lacking awareness of what the recipe entails or what steps remain.  &
``Alexa maybe could give me the bigger picture in the introduction. `\emph{This is basically what we are going to do, and let me guide you through, step by step.}'\shortspace'' (C10) \\

2 & Information Overload    &
Too much information is provided by the voice assistant at once. &
\emph{C3 requested that this step be repeated twice}: ``Step 2. Heat a large skillet over medium-high heat. Add the ground beef, breaking up with a spoon. Cook until browned, about five to seven minutes. Drain off excess grease. When you're ready, say `repeat' or `next step.'\shortspace'' \\

3 & Fragmentation           &
Information is broken up in a way that makes it difficult to act upon. &
``All of the garnishing, {[}Alexa{]} told me them in three different steps \ldots So I had to stand there, wait for it to say it, and then be like, 'Alexa, next step, previous step, previous step.'\shortspace'' (C9) \\

4 & Time Insensitivity       &
Time-sensitive directions are delivered after they are needed.  &
``A lot of information that I think is really important is time\ldots{} 
like preheating the oven \ldots{}If you're not preparing [preheating or thawing] that ahead of time, then either you're gonna be waiting 20 minutes or you'll just be going ahead with whatever, cold meat or something which will cook slower.'' (C1) \\

5 & Missing Details         &
Useful details are left out of the recipe. &
\emph{C11 requested the bolded text be added to the instructions.} ``In the same pan, heat the remaining 1 Tbsp olive oil \textbf{on medium heat}\ldots{} Cook until the juices run clear. \textbf{If chicken is not consistent thickness, consider cutting into chunks.}''
\\

6 & Discarded Context       &
Answers are based on external resources instead of the recipe. &
\emph{C2}: Alexa, how much vinegar do I need? \newline \emph{Alexa}: \alexa{From ``reference.com,'' most medical studies suggest taking no more than two tablespoons of vinegar before a meal.} \\

7 & Failure to Listen     &
The voice assistant does not respond to requests or interruptions. &
``Alexa, what should I do with the sausage? {[}\emph{11 seconds pass while C8 chops onions}{]}\ldots{} She ignored me.'' \\

8 & Uncommunicated Affordances        &
The voice assistant is not clear about its features. &
``You cannot expect it to answer any questions you ask. You need to think, `Okay, I have this problem, and in what way it can assist me.'\shortspace'' (C7) \\

9 & Limitations of Audio    &
Desiring visual information or affordances. &
``It would also be nice to have a visual image of what the sauce is supposed to look like, or the chicken.'' (C4) \\ \bottomrule

\end{tabular}
\Description[Nine key challenges]{Nine key challenges corresponding to our findings in the Results section accompanied by descriptions and representative observations from the video transcripts.}
\caption{Challenges arising from instructions dictated in audio-only format.} 
\label{tab:challenges}
\end{table*}

\begin{figure*}[]
    \centering
    \includegraphics[width=\textwidth]{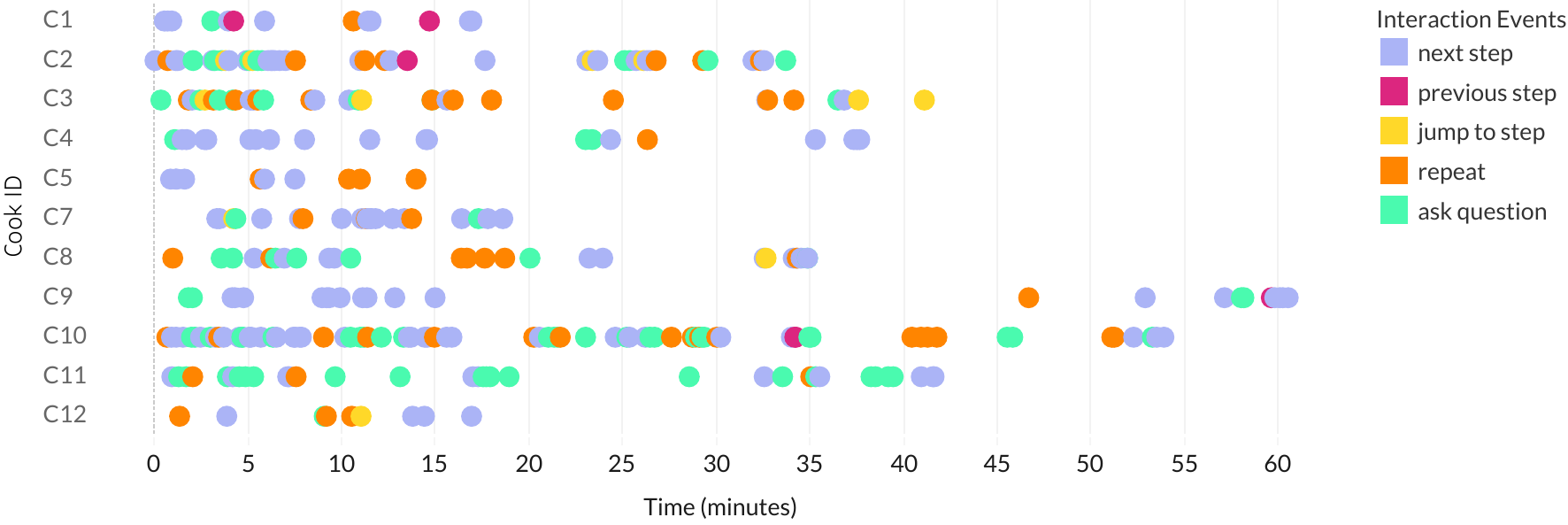}
    \Description[Timelines for each cooking session]{12 timelines representing the next step, previous step, jump to step, repeat, and ask question interactions in a cooking session.}
    \caption{Timelines of cooking sessions with Alexa. 
    \textmd{Dots represent requests that cooks made to Alexa; colors of dots represent the type of request they made. These timelines reveal a difference between navigation and information-seeking while using a voice assistant: although cooks usually navigated through recipes by requesting the next step, rarely going backwards or jumping directly to a specific step, they differed widely in how often they asked for additional information. C11 in particular asked many questions throughout her session while C5 asked none.}
    \textmd{10 events are omitted from the records of C8 and C9 due to transcription issues.}}
\label{fig:timelines}
\end{figure*}

This section presents an overview of the participants' interactions with Alexa, followed by 9 key challenges  they encountered while being guided through their recipes (see Table~\ref{tab:challenges}). To ground our results in the task at hand, we refer to participants as ``cooks'' with pseudonyms C1--12.\footnote{Video and audio data from C6's session are omitted due to technical issues.} Quotes from participants are sometimes lightly edited for brevity and clarity.

\subsection{Overview}
\label{sec:overview}

Cooks followed recipes ranging in familiarity, complexity, length, and cultural origin, adding to the richness of their experiences beyond self-reported cooking skill and frequency of using voice assistants. Most recipes were entr\'ees, with two being baked goods (C2, C8) (see Figure~\ref{fig:food}). Of the 12 cooks, 6 reported being unfamiliar with their recipe, 1 reported moderate familiarity, and 4 reported being familiar. Cooks followed an average of 8 steps per recipe ($\sigma$~=~2.9, min~=~3, max~=~11). Recipes may have included more than this number of steps: as noted in Section~\ref{sec:missing_picture}, cooks sometimes ended their sessions too early because they were unaware that there were additional steps that Alexa had not yet read. Cooking sessions ranged from 15 minutes to just over an hour.

Cooks typically interacted with Alexa dozens of times while completing their recipes. While cooks shared similar patterns of navigating through recipes, they diverged in patterns of information-seeking. The most common request was to advance to the next step or ingredient, with cooks requesting to advance an average of 13.6 times per session ($\sigma$~=~7.8, min~=~4, max~=~31). Cooks moved backwards or jumped from step to step less often: they requested to backtrack an average of 0.5 times per session ($\sigma$~=~0.8) and jumped to a specific instruction an average of 5.3 times ($\sigma$~=~5.0). Cooks frequently asked Alexa to repeat itself---4 times on average---with some asking more often ($\sigma$~=~5.0). As we discuss in Section~\ref{sec:info_overload}, frequently requesting Alexa to repeat a step usually implied that cooks were feeling overwhelmed by the amount of information they were receiving. Cooks asked many questions on average ($\mu$~=~7.4) but varied more widely on this than with any other request ($\sigma$~=~8.3). Requests of all kinds occurred throughout the session, rather than exclusively at the beginning or end (see the timelines in Figure~\ref{fig:timelines}).

Before and after each session, we asked cooks to report on a 5-point Likert scale how likely they would be to use a voice assistant to follow a recipe in the future. Most (7 of 12) cooks responded to this item in the pre-study questionnaire, yielding a median of 4 out of 5 ($\sigma$~=~0.9). These same cooks reported a half-point drop in willingness after their sessions (median~=~3.5, $\sigma$~=~1.2). When including all cooks who responded at the end of their session (12 of 12), their median willingness dropped a half point more (median~=~3, $\sigma$~=~0.97). Only one cook (C5) reported being more willing to use a voice assistant for recipe guidance after completing the study. These drops may have been influenced by the challenges described in the following sections.

\subsection{Missing the Big Picture}\label{sec:missing_picture}

When following a recipe with Alexa, cooks often felt they were missing the ``big picture.'' With a conventional written recipe, cooks can skim it beforehand to familiarize themselves with the steps they need to follow and the order they need to be followed in. With Alexa, there was no comparable way to skim. Cooks could listen to the recipe as a whole before beginning to follow it, but very few chose to do so because it is time-consuming. This led to an experience where, as C8 described, ``everything was a suprise.''

Because of this, one of the most requested features was the ability to get an overview of a recipe. One cook called this the ``bigger picture'' (C10). Five cooks explicitly mentioned that they would have liked some kind of overview of the contents of a recipe. Most of these cooks envisioned a summary that could be stated at the beginning of the recipe. C9, for instance, sketched out a summary she would have wanted to hear, which included a list of equipment, a distillation of the 11 steps into 3 ``major steps'' divided by wet and dry ingredients, and a list of preparatory steps to be performed in advance of the recipe. C12 described an overview as consisting of about a sentence per step, and C1 desired the ability to ``scroll through the whole recipe'' ahead of time.

Particularly long or involved steps could have used overviews of their own. C1, for instance, described how he would have liked a brief description of a step, and the step after, before starting the current step:

\begin{quote}
    If there was a concise summary, like, ``Add the cheese, add the pepper, cream the butter again. Next you will be preheating the pan and turning the oven on.'' A little snapshot of where you're going next. What the next turn is in the directions.
\end{quote}

Cooks sometimes wanted a better sense of how far they had progressed in the recipe. C3 described this as understanding ``where you are in the context of all the steps.''
Without a clear sense of progress, some cooks were confused when they eventually reached the end of the recipe. After Alexa narrated the final step and became silent in his session, C10 exclaimed, ``So that's it? Alexa, is that the end of the recipe?'' C10 was not the only one to experience this confusion: C3 asked for what she called a ``The End message'' because she ``wasn't sure if the last step was the last step.''
Even more crucially, five of eleven cooks missed the last step without knowing there was more to be heard. Luckily, in these few cases, the final step was either able to be inferred or had little consequence toward completing the recipe, but this information still appeared in the original script and they were not made aware of it.

\subsection{Information Overload}
\label{sec:info_overload}

When instructions appear in print, cooks have control over how much to read and when. When instructions are delivered by audio, this control is considerably diminished. A voice assistant necessarily makes decisions about how much of the instructions to read at a time, and these decisions may be poorly calibrated to users. Alexa's approach was to read instructions one step at a time. These steps were defined by the authors of the original recipe, so they they varied a great deal in their complexity. While many steps were short, simple, and memorable, others called for cooks to perform many disparate actions, making use of many ingredients.

This led a few cooks to explicitly describe their preference to hear simpler instructions. C3 wished for shorter steps, rather than ``multiple sentences within a step and having to repeat.'' C5 similarly desired that Alexa could ``[break a longer step] down into steps like the same way they do recipe ingredients.'' By default, Alexa read the ingredients in pairs. It sometimes segmented long steps into a couple of sentences, but this did not seem to be enough.

Beyond these two cooks, many seemed to struggle with remembering the instructions that were read aloud. One indication that the instructions were too long to remember is that cooks frequently asked Alexa to repeat instructions. Every cook asked Alexa to repeat at least one step. Across all sessions, cooks requested that Alexa repeat a step an average of 4 times per session. Given that recipes had an average of 8 steps, a sizeable portion of requests were repetitions. Repetitions were requested for 26 different steps, with cooks requesting at least two repetitions for 13 steps and at least three for 5 steps. Some of these steps contained quite a few details, like one that C10 asked to have repeated:

\begin{quote}
\alexa{Step 7. Put the avocados in a large bowl and gently toss with the tomatoes, lemon juice, shallots, two tablespoons oil, half teaspoon salt, and the reserved herbs. Transfer to a serving bowl.}
\end{quote}

This instruction refers to seven ingredients (two of them with accompanying measurements), two pieces of equipment (large bowl and serving bowl), and three separate actions (putting, tossing, and transferring). We can understand why, when read all at once, a step like this requires repetition: it contains many individual details, some of which must be recalled precisely.

Post-hoc analysis of the repeated steps suggests that cooks were more likely to request repetitions for steps that were more complex. We observed a correlation between the number of repetitions and various aspects of complexity of an instruction, including the number actions a cook was asked to perform, the number of ingredients they needed to use, and the number of words and sentences all in a single step. On average, steps that were repeated had 1.1 additional actions, 1.2 additional ingredients, 10 additional words, and 0.7 additional sentences compared to those that were not repeated.

Another indication that the steps were too complex is that cooks explicitly indicated many steps they would have liked to split up in their annotations of printed recipes. Six cooks indicated at least one step that they wished had been further divided. Some of these steps were truly immense, representing many individual actions, like this step that C3 would have liked to split into six component steps (as indicated by ``\big/''):

\begin{quote}
Pound the pork chops with a meat mallet or a heavy skillet until about 1/4 inch thick: \big/
season with salt and pepper. \big/
Put the flour in a shallow baking dish. \big/
Whisk the eggs, 1/2 teaspoon sesame oil and a pinch each of salt and pepper in a second dish. \big/
Put the panko in a third dish. \big/
Working with 1 chop at a time, coat in the flour and then dip in the egg, shaking off any excess; firmly press both sides in the panko.
\end{quote}

\noindent A voice assistant cannot always control how recipes are written, so it may need to guide users through instructions like the one above. It can, however, control how it processes information before delivering it. Voice assistants should play an active role in chunking information into steps that are easier for users to follow.

Many sources of information overload came from too many steps being presented at the same time, but Alexa sometimes described a single step in too much detail. Some instructions stated the obvious: C5 indicated that he did not need Alexa to tell him to ``place salmon in an ovenproof pan'' before baking it, maybe because he could have inferred it from context. In other cases, the steps included tips that the cook felt they did not need. C6, for example, reported that she would have preferred skipping suggestions for washing clams and mussels; C3 wished Alexa had omitted a suggestion to test if a pan of oil had heated up enough by tossing a breadcrumb in and watching it sizzle. Five cooks annotated printed copies of their recipes in a way that suggested information should have been left out. Along with splitting and chunking complex steps, voice assistants can identify extraneous information---which may depend on individual preferences---and omit it from the instructions altogether.

\subsection{Fragmentation}
\label{sec:fragmentation}

Delivering details in the right place is especially crucial when a recipe cannot be read but only heard. In our study, information was often fragmented across the recipe. A cook reading a recipe can search for information across the page at their own pace. A cook listening to one is dependent on the voice assistant to do the same.

Information about ingredients was particularly fragmented in our study: the amount of each ingredient to use was often present in the ingredients list but missing in the step that used it. In one case, C4's recipe called for ``1 tablespoon butter,'' ``1 tablespoon garlic,'' and ``1 tablespoon ginger,'' but the step that combined them simply said to ``melt the butter and add the garlic and ginger.'' This was already the sixth step of the recipe, so C4 had probably forgotten the quantities of the ingredients---in fact, she asked Alexa to tell her some of them at this point. In total, seven of eleven cooks explicitly asked for the amount of an ingredient at some point while cooking.

Recipes are often written in this style, perhaps to save room on a page, but details are lost when voice assistants directly read the instructions without redistributing key information. This also leads to a variety of challenges because cooks may follow the same instructions in different ways: some portioned out the ingredients as they heard them in the beginning, while others completely skipped the ingredients list because they assumed, incorrectly, that Alexa would give them all the information they needed later. 

Without awareness of steps to come, ambiguity about the number of ingredients sometimes led to deviations from the recipe. C9's recipe called for onions in two different steps. Unaware of this fact, she used all of her onions in the first step that called for them. When she arrived at the next step, she had no more onions to use.

Beyond providing all relevant details when they are needed, cooks also felt that some instructions could be combined to make them easier to perform. C9's recipe asked her to prepare the pan and fry an onion, then add cashew nuts and raisins, and finally add whole clove, cardamom pods, bay leaf, and cinnamon stick. She saw these three consecutive steps in her recipe as substeps of the same action, so she wished that they had been described together. In another session, C4 wished that ingredients in her recipe were grouped by the part of the dish they were used to make (i.e., seasoning, sauce, fried chicken, and toppings). Voice assistants will need to act as users' eyes when reading instructions for them, picking up information that has been fragmented across the page and presenting it together at the right time.

\subsection{Time Insensitivity}
\label{sec:time_insensitivity}

Recipes are often full of time-sensitive information that, if not properly anticipated, lead to problems with completing them correctly and on time. With written recipes, cooks commonly scan the text to discover and plan for such time-sensitive steps. However, cooks in eyes-free settings do not have the same luxury. Some cooks in our study found themselves in awkward situations where they could have benefited from information that was delivered too late.

In some cases, cooks ended up wasting time that was supposed to be used to complete multiple tasks in parallel. Alexa told C2, for instance, to ``\alexa{cool the cakes in the pan for fifteen minutes and then turn the cakes on a rack to cool completely.}'' About an hour later, after waiting for the cakes to cool completely as instructed, C2 asked for the next step: ``\alexa{While the cakes cool, make the buttercream.}'' C2 was supposed to start the buttercream far before she had, but she had no idea without Alexa warning her or helping her preview the next step. Similarly, C3 could have started preparing a side salad while frying her pork, but she waited to finish the current step before asking for the next.

Some cooks anticipated the need to prepare for later steps but still struggled to find time-sensitive information. C1 and C8, for example, realized early on in their sessions that they would need to preheat the oven. When asking Alexa directly for the proper settings did not work, they settled for repeatedly asking for the next step until they found the information. Voice assistants can help users anticipate time-sensitive steps by surfacing them to the beginning when remixing instructions.

\subsection{Missing Details}
\label{sec:missing_details}

In addition to providing details too often or in unhelpful places, Alexa sometimes excluded information that would have helped cooks. In particular, Alexa often excluded parenthetical information from the original written recipes. One recipe read, ``In a medium bowl, whisk together All-Purpose Flour (1 1/2 cups),'' but Alexa omitted ``(1 1/2 cups)'' when reading the ``script'' aloud. Alexa already seems to ``rewrite the script'' in some ways, but this particular approach increased fragmentation. Every cook who experienced these omissions wished Alexa had not left this information out.

Other times, Alexa excluded details about how ingredients should be prepared even though the ingredients list often included this information. For example, when one ingredients list called for ``1 Tbsp Fresh Ginger, \textit{crushed},'' Alexa left out ``crushed.'' This is doubly problematic when this information does not appear anywhere else in the recipe, like in C9's case. C9's written recipe for egg biryani called for ``5 eggs, boiled,'' but Alexa excluded the word ``boiled'' when reading aloud the ingredients list. The rest of the recipe never stated to boil the eggs, either. C9 knew ahead of time to boil the eggs because she had glanced at the ingredients list online while grocery shopping: Alexa failed on all counts to inform her of the proper preparation for the eggs. Another user relying completely on Alexa may have cracked the eggs in or overcooked the rest of their dish while waiting to boil the eggs in the middle of the recipe, especially if they were less familiar with egg biryani. 

In extreme cases, omitting parts of the original recipe led to a mistake that could not be reversed. C11, for instance, was preparing a two-part dish consisting of seasoned chicken and a yogurt sauce. Alexa told her to pour the sauce over the chicken as part of the last step but skipped an author's note at the bottom of the page that suggests storing the chicken and sauce separately if saving the dish for later. C11 was dismayed when she discovered the note while annotating the printed copy at the end of the observation session. The serving size of the dish had been much larger than expected---another case of missing the bigger picture---and she had wanted to save the leftovers for another time. Two other written recipes had similar notes at the end, which Alexa did not read aloud. All cooks who discovered omitted author's notes on paper recipes expressed that they would have liked to hear them while cooking.

Cooks also noted that they would have benefited from a voice assistant adding some details beyond the original recipe. This would be especially helpful for cooks who are unfamiliar with a dish or less experienced with cooking. C11 was not familiar with the Lebanese chicken fatteh recipe she had chosen. After cooking, she annotated her recipe with additional details she would have appreciated Alexa adding while guiding her (see \textbf{boldface} text):

\begin{quote}
    In the same pan, heat the remaining 1 Tbsp olive oil \textbf{on medium heat}; add the chicken breast to the pan and season with the garlic powder, coriander, thyme, paprika, and salt and pepper to taste. Cook until the juices run clear. \textbf{If chicken is not consistent thickness, consider cutting into chunks.}
\end{quote}

\noindent These clarifications help make assumptions about how to perform tasks more explicit. Three other cooks (C3, C11, C12) annotated their recipes with similar clarifications, including whether rice should be ``al dente'' or fully cooked and whether one or both sides of a pork chop should be cooked for the stated amount of time. Voice assistants with advanced general knowledge and commonsense reasoning skills can do more than just read instructions aloud: they can make them even more informative at the same time.

Cooks frequently asked questions to uncover these hidden details, including which type of grater to use for carrots (C3), whether a regular skillet could be used instead of nonstick (C3), how to achieve a certain consistency with sauce (C4), what a reduction was (C7), what deglazing a pan meant (C8), how long it takes to cook rice \textit{al dente} (C9), and what a saucepan was (C11). These questions represent a wide range in cooking experience, which affects whether including certain information clears up the big picture or overloads users with information. Providing the right information adapted for each user and and answering questions well while delivering the recipe could make working with voice assistants even more powerful than following instructions alone.

\subsection{Failure to Listen}
\label{sec:fail_listen}

When cooks tried to interact with Alexa, it often failed to respond. Surprisingly, Alexa often failed to respond even when cooks addressed it in the recommended way: by prefixing their requests with the wake word, ``Alexa.'' In fact, it failed to respond to six cooks who addressed it with the wake word, and five cooks faced this issue at least three times. This failure to respond was a stumbling block for conversation, with cooks waiting an average of 4.6 seconds (min~=~2s, max~=~9s) before trying the same request again. This led one cook to suggest that Alexa did not ``hear properly'' (C8).

Cooks faced even more trouble when addressing Alexa without the wake word. This may seem like a user error, but the real issue may be that Alexa is not well tuned to the way cooks naturally want to address it during longer-form, multi-turn interactions. The vast majority---eight of eleven---of cooks addressed Alexa without the wake word at least once. These failed requests caused a delay as well, with cooks waiting an average of 3.5 seconds (min~=~1s, max~=~5s, outlier~=~28s) before repeating the request with the wake word. Cooks may have been particularly confused because Alexa actually did not always require the wake word at all. Rather, after dictating a step, it would ``listen'' for follow-up requests for a few seconds before turning the microphone off. During this period, users can interact with Alexa without the wake word, but it was not always obvious that Alexa was ready for new requests. C8 even remarked, ``I don't know where or when [to use the wake word] so I just call her name every time.'' Alexa did light up while listening, but visual cues may be invisible or unclear during eyes-free interaction.

Alexa failed to respond to a wide variety of requests, including continue, repeat, next, start over, and answer a question, many of which were basic navigation requests that were unambiguously intended for the voice assistant. Features for interactions without wake words, like ``conversation mode'' on the new Amazon Alexa Echo Show 10, may help reduce the friction of following instructions with a voice assistant if it can clearly indicate when it is listening and successfully respond when prompted.

Another source of friction in cooks' conversations with Alexa was its failure to respond to interruptions. Overall, cooks seemed hesitant to interrupt Alexa, even during a very long step. The few times a cook did try to interrupt it, Alexa did not always respond. A small minority of three cooks tried to interrupt Alexa at all during their sessions, most often to stop it in the middle of providing an obviously irrelevant answer to a question they had asked (see Section~\ref{sec:discarded_context}). This was particularly frustrating when Alexa's answers were verbose. In the words of C11, ``When I asked her a question and she just, like, launched into a thing, I felt like there was no way to get her to stop. Maybe I'm just not loud enough, or maybe I didn't try, but [Alexa's answer] was just very, very long. And I was like, `All right, I just have to ride this out.'\shortspace''

\subsection{Discarded Context}
\label{sec:discarded_context}

Perplexingly, Alexa seemed to forget about the recipe when cooks asked it a clarification question. It seemed to answer questions by searching for external references instead, like in this exchange:

\begin{quote}
  \textbf{C3}: Alexa, how much flour? \\
  \textbf{Alexa}: \alexa{From cooksinfo.com, if you are making bread in North America, some suggest three parts all purpose flour to one part cake flour\ldots{}}
\end{quote}

This exchange is representative of a pattern we observed for five cooks (twelve times total), in which a cook asked a question that they wished to have answered from the recipe, but Alexa answered by referring to an external reference (i.e., ``\alexa{From \emph{<URL>}...}'', ``\alexa{According to an Alexa Answers contributor...}''). Even more perplexingly, Alexa occasionally provided a response that sounded plausible without saying that it was from an external source:

\begin{quote}
  \textbf{C1}: Alexa, how long do I need to bake the cake? \\
  \textbf{Alexa}: \alexa{It is recommended to bake cake in the oven for 30 minutes to 35 minutes at 350 degrees Fahrenheit and to rest for 10 minutes. Bake until cake is golden. Test the cake by inserting a toothpick. If the toothpick comes out clean, your cake is done.}
\end{quote}

This exchange clearly deviates from the original recipe. It calls for the oven to be heated to 350\textdegree{}F (rather than 325\textdegree{}F, as instructed by the original recipe), recommends cooking for 30--35 minutes (rather than 35--40 minutes), and suggests baking until the cake is golden (the participant was baking a red velvet cake, which would not have turned golden no matter how long it was cooked). Neglecting to indicate that it is deviating from the recipe is especially risky when the response sounds plausible. Cooks who experienced this issue were asking about a variety of details from the original recipe, including quantities of ingredients, types of ingredients, substitutions, baking temperature, seasoning options, and planning (i.e., when to start preheating an oven). Future voice assistants could default to answering questions by extracting information from the source recipe before turning to external sources.

Regardless of context, Alexa struggled to answer questions in general. Of the 71 questions asked across sessions, two thirds received answers that we believe were obviously unsuitable. Along with mistakenly turning to external resources, Alexa sometimes misunderstood the intent of a cook's question, like when it attempted to set a reminder when C10 asked it to ``remind'' him of how many tomatoes he needed. Greater awareness of the kinds of questions users tend to ask during the task completion process can help voice assistants answer them more helpfully.

\subsection{Uncommunicated Affordances}
\label{sec:uncommunicated_affordances}

Even though best practices in human-AI interaction recommend that AI-infused interfaces be clear about what they can do and how well they can do it~\citep{amershi_guidelines_2019}, Alexa did not seem designed to make its affordances for following instructions easily understood. Cooks learned that Alexa, in contrast to a human partner, required a certain way of making requests and asking for help. C7 initially communicated with Alexa as though it were a ``real person'' but ended up with a much more restricted perception of Alexa's capabilities later on:

\begin{quote}
    I have to know how it processes my information, like, to talk with it as it can understand\ldots{}
    Sometimes, instead of asking direct questions, I may ask it to repeat the instructions and figure it out myself. 
\end{quote}

Despite Alexa's efforts to communicate its numerous affordances for helping people follow step-by-step instructions, cooks were still unaware of many of them. This may have happened because Alexa did not inform the cooks of its relevant affordances at the right time. Furthermore, communicating affordances in an eyes-free setting is not as simple as ambiently displaying them on a screen for users to discover on their own. When cooks were unaware of an affordance, they usually worked around it instead of experimenting through trial and error or trying to find it through documentation. In one instance, C4 did not try to ask Alexa questions about the recipe that required external information because she ``thought that Alexa could only tell [her] what was in the recipe.'' The discoverability of affordances on audio-first interfaces may rely on efficiently informing users right when they need them.

Another affordance that was not communicated well to users was Alexa's behavior when reading ingredients. Rather than reading them all at once, Alexa read them in pairs with pauses in between. It advanced to the next pair of ingredients when requested. Two cooks were unaware of this behavior, which led to two different issues. C11 skipped the ingredients list altogether, anticipating that Alexa would read ``a whole long list'' without pauses. This made matters more difficult for her because she had to collect ingredients later as they were mentioned in the recipe. C9, in contrast, tried to listen to the ingredients list. When Alexa paused after the first two ingredients, she seemed surprised and asked, ``Is that all the ingredients it's gonna give me?''

Complicating matters, Alexa gave mixed signals about the availability of affordances, perhaps because of speech recognition issues. For instance, C8 asked Alexa for the instructions at the beginning of a recipe. Alexa responded with the statement, ``That command is not supported right now,'' even though it does in fact have this ability (and several other cooks used it successfully after a similar request). This cook understandably responded by asking, ``Oh no, do I have to start again?'' and then searched through Alexa's recipe library for the same recipe to start over instead of retrying the request. Providing clearer error messages and suggestions for working through them could have helped C8 recover from this error, as recommended in prior research~\citep{langevin_heuristic_2021}. Altogether, these cooks' experiences revealed that affordances will likely need to be more explicitly communicated by future voice assistants.

\begin{figure*}[]
    \centering
    \includegraphics[width=\textwidth]{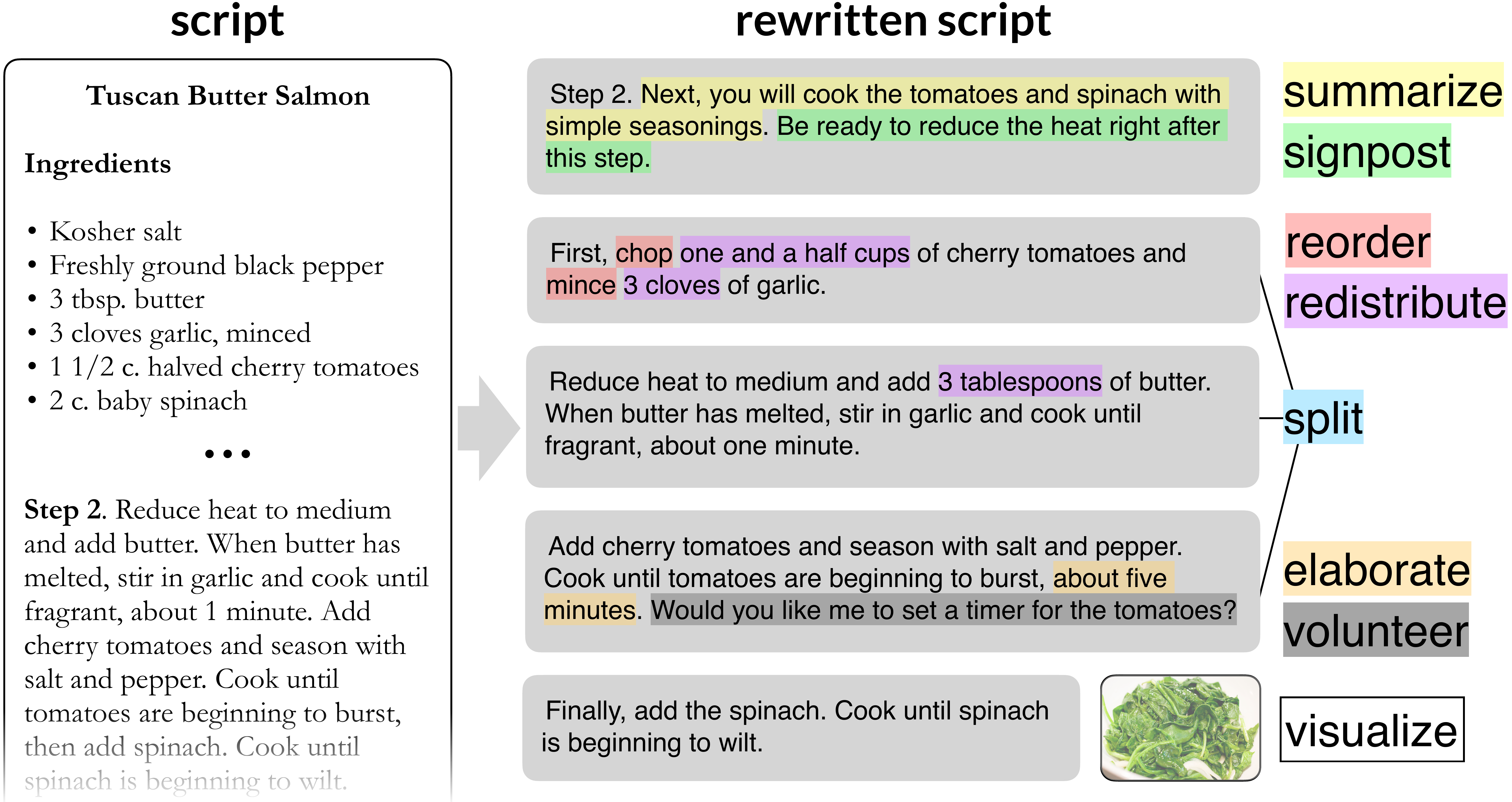}
    \Description[Rewriting the script]{A breakdown of how a script can be rewritten into easy-to-follow instructions with our suggestions from the Discussion.}
    \caption{A sample transformation after ``rewriting the script.'' \textmd{Among other changes, a set of instructions should be \emph{split} into easier-to-follow chunks; information should be \emph{redistributed}, with details appearing where they would be most useful; and the voice assistant should \emph{summarize} and \emph{signpost} to help users understand where they are in a procedure. An effective voice assistant for providing instructions will have to perform all of these tasks in a coordinated way to effectively provide task support. Example from Tuscan Butter Salmon recipe~\citep{miyashiro_tuscan_2023}. Photo of spinach by Jessica and Lon Binder on Flickr~\citep{binder_perfectly_2009}.}}
    \label{fig:rewrite}
\end{figure*}

\subsection{Limitations of Audio}
\label{sec:lim_of_audio}

We purposefully used an audio-only device for our study because we wanted to learn more about how voice assistants of all kinds can better communicate with their users. Although audio-only guidance has shown great promise in our observations, cooks sometimes wanted visual information as well.

Cooks wanted visual information to help them assess if they had achieved the intended outcome of a step, like the proper consistency of cake batter (C2) or the doneness of fried chicken (C4). Cooks also described a number of situations that could be answered with visual information, like what size equipment to use (C10), how to execute a technique in a recipe (C8, C12), how finely to chop an ingredient (C2), or the proportions of different ingredients in a mixture (C7). Some voice assistants can deliver this visual information through images or videos, but visual information does not necessarily need to be provided through visual output modalities. Verbally describing visual elements---like cooking chicken until it is no longer pink---can help voice assistants of all kinds communicate more effectively.

Beyond information to help visualize their tasks, cooks sometimes wished they had the ability to skim through their recipes. Skimming written recipes would have been useful to plan for upcoming steps (C11) or quickly recall details scattered throughout the instructions (C1, C4). Providing similarly efficient ways of ``skimming'' through audio-first content is not as obvious as delivering more content at a faster pace, at the risk of information overload. Displaying the instructions on a screen for users to scroll through should not be the final solution either, at least for eyes- or hands-free settings. Cooks who wanted to skim through recipes usually verbalized this as \textit{reading}, but the core of their request may be quickly absorbing information in some way, not necessarily using their eyes to do it. Out of all the challenges we discuss, this limitation of audio may require the most creativity to address.

\vspace{-.5ex}
\section{Discussion}
\label{sec:discussion}
In this section, we propose eight ways in which voice assistants can ``rewrite the script'' to transform written sources into more usable voice-based instructions. We conclude by considering the future role of voice assistants and relevant advances in natural language processing research for complex task guidance.

\subsection{Voice Assistants as Rewriters of Scripts}
\label{sec:recs}
We propose eight key capabilities that a ``rewriter of scripts'' should have, which are grounded in our observational study (Sec.~\ref{sec:results}). We believe these capabilities are especially suited to the current era of computing given the recent advances in natural language processing research (see Table~\ref{tab:hci_nlp_refs}). We offer a concrete vision of what rewriting the script might look like in Figure~\ref{fig:rewrite}, which includes: 

\textbf{Summarize.} Because listening does not currently afford skimming as easily as reading, voice assistants should help users familiarize themselves with the instructions by providing overviews at different levels (Sec.~\ref{sec:missing_picture}). Summarizing instructions as a whole, and particularly complex steps within them, would help users develop a sense of what the instructions entail and how to prepare for upcoming tasks. Furthermore, advanced users could use these summaries instead of the original steps if they do not need detailed guidance.

\textbf{Signpost.} Voice assistants can also provide more direct guidance by signposting. Contrary to a summary, a signpost tells the user where specific information is or what commands they can use. {Telling a user, for instance, that they are on ``step 2 \textit{of 12}'' as opposed to just ``step 2'' can help them keep track of their progress within the big picture (Sec.~{\ref{sec:missing_picture}}). Alerting users of time-sensitive steps like preheating an oven can help them anticipate actions that need to be executed in parallel (Sec.~{\ref{sec:time_insensitivity}}). Finally, simply telling users what they can say to their voice assistant would go a long way in communicating affordances (Sec.~{\ref{sec:uncommunicated_affordances}}).}

\textbf{Split.} To avoid burdening users with information overload, voice assistants can reduce the amount of information in each step of the procedure. Our study implied that simpler steps---containing fewer actions, materials, words, and sentences---were less likely to be repeated by users (Sec.~\ref{sec:info_overload}). As a preliminary rule of thumb, we suggest splitting complex steps so that each step contains one main action. Additional actions within the same step should be small or tightly related to it. Voice assistants may need to insert more pauses when an instructions contains many parts, such as many different materials or additional implied substeps.

\textbf{Elaborate.} Sometimes, voice assistants need to elaborate on small details. Voice assistants should also ensure that they do not omit important information (like crucial details in parentheses; Sec.~\ref{sec:missing_details}). Some cooks in our study appreciated when implicit details were made explicit, like that tomatoes should be cooked until they are \textit{beginning to burst}. Voice assistants should anticipate when additional details would benefit particular users, preferences, or levels of experience and provide them while delivering the instructions.

\textbf{Volunteer.} Cooks in our study sometimes implied that they would appreciate more proactive voice assistants as opposed to strictly reactive ones. Proactively volunteering information can help users anticipate the currently uncommunicated affordances of voice assistants (Sec. \ref{sec:uncommunicated_affordances}). Voice assistants can continue offering information about its affordances directly after a relevant interaction---i.e., telling a user they can say ``repeat'' to hear the current step again---and dive deeper into the content of the instructions---i.e., volunteering to set a timer or elaborate on an obscure technique.

\textbf{Reorder.} Order matters in instructions. It is especially important to for time-sensitive tasks (Sec.~\ref{sec:time_insensitivity}). Instructions that depend on each other should be detected and stated far enough in advance that users can act upon them before too late. This may require splitting steps into multiple substeps or even alerting the user well before they begin the main part of the instructions---so they can thaw frozen ingredients before cooking, for example.

\textbf{Redistribute.} When information is fragmented across a written procedure, voice assistants should group it back together. {Information in our study was particularly fragmented across the ingredients list and main instructions (Sec.~{\ref{sec:fragmentation}}). The ingredients list often contained crucial information about the amount and preparation of an ingredient (``3 cloves garlic, minced'') without repeating it when it was needed (``When butter has melted, stir in garlic and cook until fragrant, about 1 minute.'').\footnote{Examples from Tuscan Butter Salmon recipe \citep{miyashiro_tuscan_2023}.}} Redistributing this information, even if that means repeating it, would help users access information in a modality that is hard to search through.

\textbf{Visualize.} {Visual information can help offset some of the limitations of audio (Sec.~{\ref{sec:lim_of_audio}}).} Voice assistants can provide this information in two ways. First and foremost, voice assistants should verbalize visual representations, like by suggesting the user to ``cut potatoes into slices as thick as a pencil.'' Multi-modal voice assistants can display an analogous visual on their screens after generating or querying for it. These multi-modal assistants should still take care to verbalize visual information because the screen is meant to complement voice interaction, not replace it.

\subsection{Limitations}
Our conclusions are limited in several ways. First, the challenges we identified may not represent the full range of experiences of a broader population. The participants in our study were primarily college-educated, English-speaking young adults who were likely already aware of voice assistant technology. Second, our findings may not apply to all voice assistants since we used one type of device. Finally, the challenges associated with recipes may manifest differently in different types of instructions. Recipes tend to include many actions and materials (i.e., ingredients) in a single step, so our study may overrepresent issues of information overload. Examples of fragmentation related to ingredients lists are likely unique to recipes as well. Furthermore, recipes have lower stakes compared to safety-critical procedures like driving and surgery. Because taste is subjective, the outcome of a recipe is also more flexible, unlike building furniture or submitting legal documents.

\subsection{Future Work}
\label{sec:future_work}

\begin{table*}
\renewcommand{\arraystretch}{1.5}
\begin{tabular}[]{ll>{\raggedright}p{5cm}>{\raggedright}p{4.5cm}>{\raggedright}p{4cm}}

\toprule
\# & HCI Goal & Description & Related NLP Tasks & Selected Research \tabularnewline \midrule
& \textbf{Rewrite}   & Adapt written instructions into a form more easily consumed over audio. & Task-Oriented Dialogue, Text Simplification,  Style Transfer               & 
\citet{budzianowski_multiwoz_2018} \newline
\citet{reif_recipe_2022} \newline
\citet{wu_tod-bert_2020} \newline
\citet{zhang_small_2020} \tabularnewline

\arrayrulecolor{gray}\midrule\arrayrulecolor{black}

1 & Summarize & Provide overviews of entire procedures and complex steps. & Summarization (especially for procedural text)                              & 
\citet{gao_summarizing_2022} \newline
\citet{zhong_unsupervised_2022} \tabularnewline

2 & Signpost  & Convey a user's progress and how to navigate to desired information. & Information Extraction, Event Reasoning & \citet{dalvi_everything_2019} \tabularnewline

3 & Split & Segment complex steps into easy-to-follow substeps. & Procedural Text, Event Reasoning               &
\citet{kim_bisect_2021} \newline
\citet{lyu_goal-oriented_2021} \newline \citet{zhang_automatic_2022} \newline \citet{zhou_show_2022} \tabularnewline

4 & Elaborate & Anticipate details the user wants without requiring them to ask. & Information Extraction, Commonsense Reasoning & \citet{druck_spice_2012} \newline
\citet{zhang_automatic_2022} \tabularnewline

5 & Volunteer & Proactively tell the user what affordances are available. & Question Generation                                                        & \citet{tu_competence-based_2022} \tabularnewline

6 & Reorder & Move time-sensitive steps to the point in instructions where users should begin to follow them. & Event Reasoning, Event Duration Prediction, Goal-Step Reasoning, Temporal Ordering   & \citet{kiddon_mise_2015} \newline \citet{zhang_reasoning_2020} \tabularnewline

7 & Redistribute & Repeat information that was fragmented in the written instructions whenever it is needed over audio. & Hierarchical Event Reasoning, Semantic Similarity, Relational Knowledge & 
\citet{chandrasekaran_evolution_2021} \newline
\citet{speer_conceptnet_2017}  \tabularnewline

8 & Visualize & Describe or show visual information to clarify techniques, materials, and intended results. & Visual Goal-Step Inference, Text-to-Image Generation                       & 
\citet{ramesh_zero-shot_2021} \newline
\citet{rombach_high-resolution_2022} \newline
\citet{yang_induce_2021} \newline
\citet{yang_visual_2021} \tabularnewline

\bottomrule
\end{tabular}
\Description[HCI goals and NLP tasks]{Each suggestion for rewriting the script from the Discussion along with analogous NLP tasks and selected relevant work.}
\caption{HCI goals and relevant work from NLP. \textmd{We reference relevant work in Natural Language Processing that can help the Human-Computer Interaction research community achieve the 8 goals we describe in Section~\ref{sec:recs}.}}

\label{tab:hci_nlp_refs}
\end{table*}

Our conclusions suggest directions in which the fields of human-computer interaction (HCI) and natural language processing (NLP) can together provide more effective guidance for complex tasks.

Within HCI, additional studies can further clarify what it means for a voice assistant to effectively rewrite the script by replicating our \textit{in situ} methods with other voice assistants and types of instructions. These studies should take care to include participants who represent a greater range of ability status, language proficiency, cultural origin, and age. Wizard-of-Oz studies would be especially informative for testing aspirational variants of voice assistants that can execute our suggestions as well. We also recognize that voice interaction can go beyond voice itself. Future studies can clarify the role of external displays and augmented reality in showing effective visuals at the right times to complement audio-first instructions.

In a cyclical fashion, our findings resonate with and can further inspire research efforts within NLP. Many of the goals we describe in Section~{\ref{sec:recs}} can already be achieved by leveraging current advances in well known NLP tasks, especially task-oriented dialogue, summarization, event reasoning, commonsense reasoning, question generation, and text-to-image generation. To help unite the two fields, we summarize relevant NLP research in Table~{\ref{tab:hci_nlp_refs}}. Our observational study method is effective for more than just identifying user needs: it can be a robust, user-centered way of evaluating NLP contributions. Bringing many of these techniques together into a single system capable of producing coherent, easy-to-follow text can help voice assistants develop to maturity.

\subsection{Futures with Voice Assistants that Rewrite the Script}
\label{sec:future}

Our work explores the design of voice assistants that guide users through complex tasks, even when the tasks are unfamiliar. Many solutions we propose for the challenges revealed by our observational study are already possible with current progress in natural language processing (NLP), as we discuss in Section~\ref{sec:future_work}. In this section, we consider what it would mean for voice assistants to be able to guide users through complex tasks as fluidly as we imagine.

In a future filled with voice assistants that are skilled at complex task guidance, we may fear that people's ability to learn new procedures will become diminished. As~\citet{eiriksdottir_procedural_2011} describe in their review of research on instruction design, concrete instructions that are easy to follow right away often lose their potency in transferring to new tasks. The NLP community has been grappling with a similar fear fueled by the recent release of \mbox{ChatGPT}, an immensely powerful language model~\citep{openai_chatgpt_2022}. One concerned researcher wrote that ChatGPT is a ``plague upon education'' and a ``threat to human intelligence and academic authority'' because of its ability to automate many writing tasks. \citet{duckworth_op-ed_2023}, on the other hand, argue that ChatGPT has the power to ``accelerate the trend toward valuing critical thinking'' because users need to carefully evaluate its output. In our view, voice assistants that make procedures easy to follow may remove the incentive to internalize those procedures, but they also raise the baseline of the procedures we are able to learn at all.

We may also fear that the advancement of voice assistant technology threatens the social benefits of instruction-following. We often learn procedures by following the guidance of other people, whether we are cooking new recipes~\citep{norrick_conversational_2011, claxton_cooking_2019}, administering CPR~\citep{birkun_pre-recorded_2018}, or tackling any number of other tasks. We also value exchanging additional insight and building relationships beyond the procedure itself. In today's digital age, instructions have become more diverse and accessible than ever, but they have also become less personal now that we have the option of going online instead of the necessity of seeking out experts in person. 

Like other digital resources, voice assistants can add to diversity and accessibility, without necessarily detracting from human life and relationships. We see the future role of voice assistants as increasing access to information rather than replacing human guidance. Whether they are guiding us quickly through complex instructions or leaving out details to help us practice procedural knowledge (e.g., \citep{graesser_intelligent_2001}), voice assistants can be designed for both learning and executing at the same time. Working with a voice assistant does not have to be a solitary activity, either: voice assistants can help us collaborate with each other (e.g., \citep{winkler_alexa_2019}). No matter how well they rewrite the script, voice assistants are still \textit{assistants}, and we have the power to choose how they assist us.

\section{Conclusion}
In this paper, we studied how voice assistants should be designed to guide users through complex instructions. Focusing on recipes as an example, we observed 12 people as they cooked at home while being guided by Amazon Alexa. This led us to nine key challenges that users face when modern voice assistant technology for complex task guidance falls short. Many challenges---like information overload, fragmentation, and time-insensitivity---arose from voice assistants reciting written recipes as though they were scripts. We propose eight ways for voice assistants to ``rewrite the script'' into a form that is easier to follow in hands- and eyes-free settings. Rewriting the script is crucial for any intelligent agent that communicates through spoken conversation, even devices that incorporate visual output. Future voice assistants can solve these problems by bringing together insights from human-computer interaction and natural language processing research, one step at a time.

\begin{acks}
We would like to thank the participants of our official and pilot studies. We are especially grateful to Liam Dugan for his invaluable suggestions throughout our work, Daphne Ippolito and Artemis Panagopoulou for their insight, and Hita Kambhamettu for her assistance on Figure~\ref{fig:timelines} on short notice. Finally, we thank the anonymous reviewers for their feedback. This material is based upon work supported by the National Science Foundation Graduate Research Fellowship under Grant No. DGE-1845298.
\end{acks}

\bibliographystyle{ACM-Reference-Format}
\bibliography{References/new_zotero}


\begin{thebibliography}{66}


\ifx \showCODEN    \undefined \def \showCODEN     #1{\unskip}     \fi
\ifx \showDOI      \undefined \def \showDOI       #1{#1}\fi
\ifx \showISBNx    \undefined \def \showISBNx     #1{\unskip}     \fi
\ifx \showISBNxiii \undefined \def \showISBNxiii  #1{\unskip}     \fi
\ifx \showISSN     \undefined \def \showISSN      #1{\unskip}     \fi
\ifx \showLCCN     \undefined \def \showLCCN      #1{\unskip}     \fi
\ifx \shownote     \undefined \def \shownote      #1{#1}          \fi
\ifx \showarticletitle \undefined \def \showarticletitle #1{#1}   \fi
\ifx \showURL      \undefined \def \showURL       {\relax}        \fi
\providecommand\bibfield[2]{#2}
\providecommand\bibinfo[2]{#2}
\providecommand\natexlab[1]{#1}
\providecommand\showeprint[2][]{arXiv:#2}

\bibitem[\protect\citeauthoryear{Abdolrahmani, Howes~Gupta, Vader, Kuber, and
  Branham}{Abdolrahmani et~al\mbox{.}}{2021}]%
        {abdolrahmani_towards_2021}
\bibfield{author}{\bibinfo{person}{Ali Abdolrahmani}, \bibinfo{person}{Maya
  Howes~Gupta}, \bibinfo{person}{Mei-Lian Vader}, \bibinfo{person}{Ravi Kuber},
  {and} \bibinfo{person}{Stacy Branham}.} \bibinfo{year}{2021}\natexlab{}.
\newblock \showarticletitle{Towards {More} {Transactional} {Voice}
  {Assistants}: {Investigating} the {Potential} for a {Multimodal}
  {Voice}-{Activated} {Indoor} {Navigation} {Assistant} for {Blind} and
  {Sighted} {Travelers}}. In \bibinfo{booktitle}{\emph{Proceedings of the 2021
  {CHI} {Conference} on {Human} {Factors} in {Computing} {Systems}}}
  \emph{(\bibinfo{series}{{CHI} '21})}. \bibinfo{publisher}{Association for
  Computing Machinery}, \bibinfo{address}{New York, NY, USA},
  \bibinfo{pages}{1--16}.
\newblock
\showISBNx{978-1-4503-8096-6}
\urldef\tempurl%
\url{https://doi.org/10.1145/3411764.3445638}
\showDOI{\tempurl}


\bibitem[\protect\citeauthoryear{Amershi, Weld, Vorvoreanu, Fourney, Nushi,
  Collisson, Suh, Iqbal, Bennett, Inkpen, Teevan, Kikin-Gil, and
  Horvitz}{Amershi et~al\mbox{.}}{2019}]%
        {amershi_guidelines_2019}
\bibfield{author}{\bibinfo{person}{Saleema Amershi}, \bibinfo{person}{Dan
  Weld}, \bibinfo{person}{Mihaela Vorvoreanu}, \bibinfo{person}{Adam Fourney},
  \bibinfo{person}{Besmira Nushi}, \bibinfo{person}{Penny Collisson},
  \bibinfo{person}{Jina Suh}, \bibinfo{person}{Shamsi Iqbal},
  \bibinfo{person}{Paul~N. Bennett}, \bibinfo{person}{Kori Inkpen},
  \bibinfo{person}{Jaime Teevan}, \bibinfo{person}{Ruth Kikin-Gil}, {and}
  \bibinfo{person}{Eric Horvitz}.} \bibinfo{year}{2019}\natexlab{}.
\newblock \showarticletitle{Guidelines for {Human}-{AI} {Interaction}}. In
  \bibinfo{booktitle}{\emph{Proceedings of the 2019 {CHI} {Conference} on
  {Human} {Factors} in {Computing} {Systems}}} \emph{(\bibinfo{series}{{CHI}
  '19})}. \bibinfo{publisher}{Association for Computing Machinery},
  \bibinfo{address}{New York, NY, USA}, \bibinfo{pages}{1--13}.
\newblock
\showISBNx{978-1-4503-5970-2}
\urldef\tempurl%
\url{https://doi.org/10.1145/3290605.3300233}
\showDOI{\tempurl}


\bibitem[\protect\citeauthoryear{Ammari, Kaye, Tsai, and Bentley}{Ammari
  et~al\mbox{.}}{2019}]%
        {ammari_music_2019}
\bibfield{author}{\bibinfo{person}{Tawfiq Ammari}, \bibinfo{person}{Jofish
  Kaye}, \bibinfo{person}{Janice~Y. Tsai}, {and} \bibinfo{person}{Frank
  Bentley}.} \bibinfo{year}{2019}\natexlab{}.
\newblock \showarticletitle{Music, {Search}, and {IoT}: {How} {People}
  ({Really}) {Use} {Voice} {Assistants}}.
\newblock \bibinfo{journal}{\emph{ACM Transactions on Computer-Human
  Interaction}} \bibinfo{volume}{26}, \bibinfo{number}{3}
  (\bibinfo{date}{April} \bibinfo{year}{2019}).
\newblock
\showISSN{1073-0516}
\urldef\tempurl%
\url{https://doi.org/10.1145/3311956}
\showDOI{\tempurl}


\bibitem[\protect\citeauthoryear{Beyer and Holtzblatt}{Beyer and
  Holtzblatt}{1997}]%
        {beyer_contextual_1997}
\bibfield{author}{\bibinfo{person}{Hugh Beyer} {and} \bibinfo{person}{Karen
  Holtzblatt}.} \bibinfo{year}{1997}\natexlab{}.
\newblock \bibinfo{booktitle}{\emph{Contextual {Design}: {Defining}
  {Customer}-{Centered} {Systems}}}.
\newblock \bibinfo{publisher}{Morgan Kaufmann Publishers},
  \bibinfo{address}{San Francisco, CA, USA}.
\newblock
\showISBNx{978-0-08-050304-2}


\bibitem[\protect\citeauthoryear{Binder and Binder}{Binder and Binder}{2009}]%
        {binder_perfectly_2009}
\bibfield{author}{\bibinfo{person}{Lon Binder} {and} \bibinfo{person}{Jessica
  Binder}.} \bibinfo{year}{2009}\natexlab{}.
\newblock \bibinfo{title}{Perfectly {Cooked} {Spinach}}.
\newblock
\newblock
\urldef\tempurl%
\url{https://flic.kr/p/61rqjw}
\showURL{%
\tempurl}


\bibitem[\protect\citeauthoryear{Birkun, Glotov, Ndjamen, Alaiye, Adeleke, and
  Samarin}{Birkun et~al\mbox{.}}{2018}]%
        {birkun_pre-recorded_2018}
\bibfield{author}{\bibinfo{person}{Alexei Birkun}, \bibinfo{person}{Maksim
  Glotov}, \bibinfo{person}{Herman~Franklin Ndjamen}, \bibinfo{person}{Esther
  Alaiye}, \bibinfo{person}{Temidara Adeleke}, {and} \bibinfo{person}{Sergey
  Samarin}.} \bibinfo{year}{2018}\natexlab{}.
\newblock \showarticletitle{Pre-{Recorded} {Instructional} {Audio} vs.
  {Dispatchers}' {Conversational} {Assistance} in {Telephone} {Cardiopulmonary}
  {Resuscitation}: {A} {Randomized} {Controlled} {Simulation} {Study}}.
\newblock \bibinfo{journal}{\emph{World Journal of Emergency Medicine}}
  \bibinfo{volume}{9}, \bibinfo{number}{3} (\bibinfo{year}{2018}),
  \bibinfo{pages}{165--171}.
\newblock
\showISSN{1920-8642}
\urldef\tempurl%
\url{https://doi.org/10.5847/wjem.j.1920-8642.2018.03.001}
\showDOI{\tempurl}


\bibitem[\protect\citeauthoryear{Blandford, Furniss, and Makri}{Blandford
  et~al\mbox{.}}{2016}]%
        {blandford_qualitative_2016}
\bibfield{author}{\bibinfo{person}{Ann Blandford}, \bibinfo{person}{Dominic
  Furniss}, {and} \bibinfo{person}{Stephann Makri}.}
  \bibinfo{year}{2016}\natexlab{}.
\newblock \bibinfo{booktitle}{\emph{Qualitative {HCI} {Research}: {Going}
  {Behind} the {Scenes}}}.
\newblock \bibinfo{publisher}{Springer International Publishing},
  \bibinfo{address}{Cham}.
\newblock
\showISBNx{978-3-031-01089-7 978-3-031-02217-3}
\urldef\tempurl%
\url{https://doi.org/10.1007/978-3-031-02217-3}
\showDOI{\tempurl}


\bibitem[\protect\citeauthoryear{Budzianowski, Wen, Tseng, Casanueva, Ultes,
  Ramadan, and Gašić}{Budzianowski et~al\mbox{.}}{2018}]%
        {budzianowski_multiwoz_2018}
\bibfield{author}{\bibinfo{person}{Paweł Budzianowski},
  \bibinfo{person}{Tsung-Hsien Wen}, \bibinfo{person}{Bo-Hsiang Tseng},
  \bibinfo{person}{Iñigo Casanueva}, \bibinfo{person}{Stefan Ultes},
  \bibinfo{person}{Osman Ramadan}, {and} \bibinfo{person}{Milica Gašić}.}
  \bibinfo{year}{2018}\natexlab{}.
\newblock \showarticletitle{{MultiWOZ} -- {A} {Large}-{Scale} {Multi}-{Domain}
  {Wizard}-of-{Oz} {Dataset} for {Task}-{Oriented} {Dialogue} {Modelling}}. In
  \bibinfo{booktitle}{\emph{Proceedings of the 2018 {Conference} on {Empirical}
  {Methods} in {Natural} {Language} {Processing}}}.
  \bibinfo{publisher}{Association for Computational Linguistics},
  \bibinfo{address}{Brussels, Belgium}, \bibinfo{pages}{5016--5026}.
\newblock
\urldef\tempurl%
\url{https://doi.org/10.18653/v1/D18-1547}
\showDOI{\tempurl}


\bibitem[\protect\citeauthoryear{Cambre, Liu, Taylor, and Kulkarni}{Cambre
  et~al\mbox{.}}{2019}]%
        {cambre_vitro_2019}
\bibfield{author}{\bibinfo{person}{Julia Cambre}, \bibinfo{person}{Ying Liu},
  \bibinfo{person}{Rebecca~E. Taylor}, {and} \bibinfo{person}{Chinmay
  Kulkarni}.} \bibinfo{year}{2019}\natexlab{}.
\newblock \showarticletitle{Vitro: {Designing} a {Voice} {Assistant} for the
  {Scientific} {Lab} {Workplace}}. In \bibinfo{booktitle}{\emph{Proceedings of
  the 2019 on {Designing} {Interactive} {Systems} {Conference}}}
  \emph{(\bibinfo{series}{{DIS} '19})}. \bibinfo{publisher}{Association for
  Computing Machinery}, \bibinfo{address}{New York, NY, USA},
  \bibinfo{pages}{1531--1542}.
\newblock
\showISBNx{978-1-4503-5850-7}
\urldef\tempurl%
\url{https://doi.org/10.1145/3322276.3322298}
\showDOI{\tempurl}


\bibitem[\protect\citeauthoryear{Carroll, Chiodo, Lin, Nidever, and
  Prathipati}{Carroll et~al\mbox{.}}{2017}]%
        {carroll_robin_2017}
\bibfield{author}{\bibinfo{person}{Clare Carroll}, \bibinfo{person}{Catherine
  Chiodo}, \bibinfo{person}{Adena~Xin Lin}, \bibinfo{person}{Meg Nidever},
  {and} \bibinfo{person}{Jayanth Prathipati}.} \bibinfo{year}{2017}\natexlab{}.
\newblock \showarticletitle{Robin: {Enabling} {Independence} {For}
  {Individuals} {With} {Cognitive} {Disabilities} {Using} {Voice} {Assistive}
  {Technology}}. In \bibinfo{booktitle}{\emph{Proceedings of the 2017 {CHI}
  {Conference} {Extended} {Abstracts} on {Human} {Factors} in {Computing}
  {Systems}}} \emph{(\bibinfo{series}{{CHI} {EA} '17})}.
  \bibinfo{publisher}{Association for Computing Machinery},
  \bibinfo{address}{New York, NY, USA}, \bibinfo{pages}{46--53}.
\newblock
\showISBNx{978-1-4503-4656-6}
\urldef\tempurl%
\url{https://doi.org/10.1145/3027063.3049266}
\showDOI{\tempurl}


\bibitem[\protect\citeauthoryear{Carroll}{Carroll}{1990}]%
        {carroll_nurnberg_1990}
\bibfield{author}{\bibinfo{person}{John~M. Carroll}.}
  \bibinfo{year}{1990}\natexlab{}.
\newblock \bibinfo{booktitle}{\emph{The {Nurnberg} {Funnel}: {Designing}
  {Minimalist} {Instruction} for {Practical} {Computer} {Skill}}}.
\newblock
\urldef\tempurl%
\url{https://mitpress.mit.edu/9780262031639/the-nurnberg-funnel/}
\showURL{%
\tempurl}


\bibitem[\protect\citeauthoryear{Chandrasekaran and Mago}{Chandrasekaran and
  Mago}{2021}]%
        {chandrasekaran_evolution_2021}
\bibfield{author}{\bibinfo{person}{Dhivya Chandrasekaran} {and}
  \bibinfo{person}{Vijay Mago}.} \bibinfo{year}{2021}\natexlab{}.
\newblock \showarticletitle{Evolution of {Semantic} {Similarity}—{A}
  {Survey}}.
\newblock \bibinfo{journal}{\emph{Comput. Surveys}} \bibinfo{volume}{54},
  \bibinfo{number}{2} (\bibinfo{date}{Feb.} \bibinfo{year}{2021}),
  \bibinfo{pages}{41:1--41:37}.
\newblock
\showISSN{0360-0300}
\urldef\tempurl%
\url{https://doi.org/10.1145/3440755}
\showDOI{\tempurl}


\bibitem[\protect\citeauthoryear{Chang, Guillain, Jung, Hare, Kim, and
  Agrawala}{Chang et~al\mbox{.}}{2018}]%
        {chang_recipescape_2018}
\bibfield{author}{\bibinfo{person}{Minsuk Chang}, \bibinfo{person}{Leonore~V.
  Guillain}, \bibinfo{person}{Hyeungshik Jung}, \bibinfo{person}{Vivian~M.
  Hare}, \bibinfo{person}{Juho Kim}, {and} \bibinfo{person}{Maneesh Agrawala}.}
  \bibinfo{year}{2018}\natexlab{}.
\newblock \showarticletitle{{RecipeScape}: {An} {Interactive} {Tool} for
  {Analyzing} {Cooking} {Instructions} at {Scale}}. In
  \bibinfo{booktitle}{\emph{Proceedings of the 2018 {CHI} {Conference} on
  {Human} {Factors} in {Computing} {Systems}}} \emph{(\bibinfo{series}{{CHI}
  '18})}. \bibinfo{publisher}{Association for Computing Machinery},
  \bibinfo{address}{New York, NY, USA}, \bibinfo{pages}{1--12}.
\newblock
\showISBNx{978-1-4503-5620-6}
\urldef\tempurl%
\url{https://doi.org/10.1145/3173574.3174025}
\showDOI{\tempurl}


\bibitem[\protect\citeauthoryear{Chang, Truong, Wang, Agrawala, and Kim}{Chang
  et~al\mbox{.}}{2019}]%
        {chang_how_2019}
\bibfield{author}{\bibinfo{person}{Minsuk Chang}, \bibinfo{person}{Anh Truong},
  \bibinfo{person}{Oliver Wang}, \bibinfo{person}{Maneesh Agrawala}, {and}
  \bibinfo{person}{Juho Kim}.} \bibinfo{year}{2019}\natexlab{}.
\newblock \showarticletitle{How to {Design} {Voice} {Based} {Navigation} for
  {How}-{To} {Videos}}. In \bibinfo{booktitle}{\emph{Proceedings of the 2019
  {CHI} {Conference} on {Human} {Factors} in {Computing} {Systems}}}
  \emph{(\bibinfo{series}{{CHI} '19})}. \bibinfo{publisher}{Association for
  Computing Machinery}, \bibinfo{address}{New York, NY, USA},
  \bibinfo{pages}{1--11}.
\newblock
\showISBNx{978-1-4503-5970-2}
\urldef\tempurl%
\url{https://doi.org/10.1145/3290605.3300931}
\showDOI{\tempurl}


\bibitem[\protect\citeauthoryear{Chen, Chi, Chu, Chen, and Huang}{Chen
  et~al\mbox{.}}{2010}]%
        {chen_smart_2010}
\bibfield{author}{\bibinfo{person}{Jen-Hao Chen}, \bibinfo{person}{Peggy Pei-Yu
  Chi}, \bibinfo{person}{Hao-Hua Chu}, \bibinfo{person}{Cheryl Chia-Hui Chen},
  {and} \bibinfo{person}{Polly Huang}.} \bibinfo{year}{2010}\natexlab{}.
\newblock \showarticletitle{A {Smart} {Kitchen} for {Nutrition}-{Aware}
  {Cooking}}.
\newblock \bibinfo{journal}{\emph{IEEE Pervasive Computing}}
  \bibinfo{volume}{9}, \bibinfo{number}{04} (\bibinfo{date}{Oct.}
  \bibinfo{year}{2010}), \bibinfo{pages}{58--65}.
\newblock
\showISSN{1558-2590}
\urldef\tempurl%
\url{https://doi.org/10.1109/MPRV.2010.75}
\showDOI{\tempurl}


\bibitem[\protect\citeauthoryear{Clark, Pantidi, Cooney, Doyle, Garaialde,
  Edwards, Spillane, Gilmartin, Murad, Munteanu, Wade, and Cowan}{Clark
  et~al\mbox{.}}{2019}]%
        {clark_what_2019}
\bibfield{author}{\bibinfo{person}{Leigh Clark}, \bibinfo{person}{Nadia
  Pantidi}, \bibinfo{person}{Orla Cooney}, \bibinfo{person}{Philip Doyle},
  \bibinfo{person}{Diego Garaialde}, \bibinfo{person}{Justin Edwards},
  \bibinfo{person}{Brendan Spillane}, \bibinfo{person}{Emer Gilmartin},
  \bibinfo{person}{Christine Murad}, \bibinfo{person}{Cosmin Munteanu},
  \bibinfo{person}{Vincent Wade}, {and} \bibinfo{person}{Benjamin~R. Cowan}.}
  \bibinfo{year}{2019}\natexlab{}.
\newblock \showarticletitle{What {Makes} a {Good} {Conversation}? {Challenges}
  in {Designing} {Truly} {Conversational} {Agents}}. In
  \bibinfo{booktitle}{\emph{Proceedings of the 2019 {CHI} {Conference} on
  {Human} {Factors} in {Computing} {Systems}}} \emph{(\bibinfo{series}{{CHI}
  '19})}. \bibinfo{publisher}{Association for Computing Machinery},
  \bibinfo{address}{New York, NY, USA}, \bibinfo{pages}{1--12}.
\newblock
\showISBNx{978-1-4503-5970-2}
\urldef\tempurl%
\url{https://doi.org/10.1145/3290605.3300705}
\showDOI{\tempurl}


\bibitem[\protect\citeauthoryear{Claxton}{Claxton}{2019}]%
        {claxton_cooking_2019}
\bibfield{author}{\bibinfo{person}{Alana Claxton}.}
  \bibinfo{year}{2019}\natexlab{}.
\newblock \emph{\bibinfo{title}{Cooking {Lessons}: {Oral} {Recipe} {Sharing} in
  the {Southern} {Kitchen}}}.
\newblock \bibinfo{thesistype}{Master's\ thesis}. \bibinfo{school}{East
  Tennessee State University}.
\newblock
\urldef\tempurl%
\url{https://dc.etsu.edu/etd/3550/}
\showURL{%
\tempurl}


\bibitem[\protect\citeauthoryear{Dalvi, Tandon, Bosselut, Yih, and Clark}{Dalvi
  et~al\mbox{.}}{2019}]%
        {dalvi_everything_2019}
\bibfield{author}{\bibinfo{person}{Bhavana Dalvi}, \bibinfo{person}{Niket
  Tandon}, \bibinfo{person}{Antoine Bosselut}, \bibinfo{person}{Wen-tau Yih},
  {and} \bibinfo{person}{Peter Clark}.} \bibinfo{year}{2019}\natexlab{}.
\newblock \showarticletitle{Everything {Happens} for a {Reason}: {Discovering}
  the {Purpose} of {Actions} in {Procedural} {Text}}. In
  \bibinfo{booktitle}{\emph{Proceedings of the 2019 {Conference} on {Empirical}
  {Methods} in {Natural} {Language} {Processing} and the 9th {International}
  {Joint} {Conference} on {Natural} {Language} {Processing}
  ({EMNLP}-{IJCNLP})}}. \bibinfo{publisher}{Association for Computational
  Linguistics}, \bibinfo{address}{Hong Kong, China},
  \bibinfo{pages}{4496--4505}.
\newblock
\urldef\tempurl%
\url{https://doi.org/10.18653/v1/D19-1457}
\showDOI{\tempurl}


\bibitem[\protect\citeauthoryear{Druck and Pang}{Druck and Pang}{2012}]%
        {druck_spice_2012}
\bibfield{author}{\bibinfo{person}{Gregory Druck} {and} \bibinfo{person}{Bo
  Pang}.} \bibinfo{year}{2012}\natexlab{}.
\newblock \showarticletitle{Spice {It} {Up}? {Mining} {Refinements} to {Online}
  {Instructions} from {User} {Generated} {Content}}. In
  \bibinfo{booktitle}{\emph{Proceedings of the 50th {Annual} {Meeting} of the
  {Association} for {Computational} {Linguistics} ({Volume} 1: {Long}
  {Papers})}}. \bibinfo{publisher}{Association for Computational Linguistics},
  \bibinfo{address}{Jeju Island, Korea}, \bibinfo{pages}{545--553}.
\newblock
\urldef\tempurl%
\url{https://aclanthology.org/P12-1057}
\showURL{%
\tempurl}


\bibitem[\protect\citeauthoryear{Duckworth and Ungar}{Duckworth and
  Ungar}{2023}]%
        {duckworth_op-ed_2023}
\bibfield{author}{\bibinfo{person}{Angela Duckworth} {and}
  \bibinfo{person}{Lyle Ungar}.} \bibinfo{year}{2023}\natexlab{}.
\newblock \showarticletitle{Op-{Ed}: {Don}'t {Ban} {Chatbots} in {Classrooms}
  — {Use} {Them} to {Change} {How} {We} {Teach}}.
\newblock \bibinfo{journal}{\emph{Yahoo Entertainment}} (\bibinfo{date}{Jan.}
  \bibinfo{year}{2023}).
\newblock
\urldef\tempurl%
\url{https://www.yahoo.com/entertainment/op-ed-dont-ban-chatbots-112037967.html}
\showURL{%
\tempurl}


\bibitem[\protect\citeauthoryear{Dunham, Lee, and Persky}{Dunham
  et~al\mbox{.}}{2020}]%
        {dunham_psychology_2020}
\bibfield{author}{\bibinfo{person}{Sabrina Dunham}, \bibinfo{person}{Edward
  Lee}, {and} \bibinfo{person}{Adam~M. Persky}.}
  \bibinfo{year}{2020}\natexlab{}.
\newblock \showarticletitle{The {Psychology} of {Following} {Instructions} and
  {Its} {Implications}}.
\newblock \bibinfo{journal}{\emph{American Journal of Pharmaceutical
  Education}} \bibinfo{volume}{84}, \bibinfo{number}{8} (\bibinfo{date}{Aug.}
  \bibinfo{year}{2020}).
\newblock
\showISSN{0002-9459}
\urldef\tempurl%
\url{https://doi.org/10.5688/ajpe7779}
\showDOI{\tempurl}


\bibitem[\protect\citeauthoryear{Eiriksdottir and Catrambone}{Eiriksdottir and
  Catrambone}{2011}]%
        {eiriksdottir_procedural_2011}
\bibfield{author}{\bibinfo{person}{Elsa Eiriksdottir} {and}
  \bibinfo{person}{Richard Catrambone}.} \bibinfo{year}{2011}\natexlab{}.
\newblock \showarticletitle{Procedural {Instructions}, {Principles}, and
  {Examples}: {How} to {Structure} {Instructions} for {Procedural} {Tasks} to
  {Enhance} {Performance}, {Learning}, and {Transfer}}.
\newblock \bibinfo{journal}{\emph{Human Factors}} \bibinfo{volume}{53},
  \bibinfo{number}{6} (\bibinfo{date}{Dec.} \bibinfo{year}{2011}),
  \bibinfo{pages}{749--770}.
\newblock
\showISSN{0018-7208}
\urldef\tempurl%
\url{https://doi.org/10.1177/0018720811419154}
\showDOI{\tempurl}


\bibitem[\protect\citeauthoryear{Fast, Chen, Mendelsohn, Bassen, and
  Bernstein}{Fast et~al\mbox{.}}{2018}]%
        {fast_iris_2018}
\bibfield{author}{\bibinfo{person}{Ethan Fast}, \bibinfo{person}{Binbin Chen},
  \bibinfo{person}{Julia Mendelsohn}, \bibinfo{person}{Jonathan Bassen}, {and}
  \bibinfo{person}{Michael~S. Bernstein}.} \bibinfo{year}{2018}\natexlab{}.
\newblock \showarticletitle{Iris: {A} {Conversational} {Agent} for {Complex}
  {Tasks}}. In \bibinfo{booktitle}{\emph{Proceedings of the 2018 {CHI}
  {Conference} on {Human} {Factors} in {Computing} {Systems}}}
  \emph{(\bibinfo{series}{{CHI} '18})}. \bibinfo{publisher}{Association for
  Computing Machinery}, \bibinfo{address}{New York, NY, USA},
  \bibinfo{pages}{1--12}.
\newblock
\showISBNx{978-1-4503-5620-6}
\urldef\tempurl%
\url{https://doi.org/10.1145/3173574.3174047}
\showDOI{\tempurl}


\bibitem[\protect\citeauthoryear{Gao, Zhang, Chen, Yan, and Zhao}{Gao
  et~al\mbox{.}}{2022}]%
        {gao_summarizing_2022}
\bibfield{author}{\bibinfo{person}{Shen Gao}, \bibinfo{person}{Haotong Zhang},
  \bibinfo{person}{Xiuying Chen}, \bibinfo{person}{Rui Yan}, {and}
  \bibinfo{person}{Dongyan Zhao}.} \bibinfo{year}{2022}\natexlab{}.
\newblock \showarticletitle{Summarizing {Procedural} {Text}: {Data} and
  {Approach}}. In \bibinfo{booktitle}{\emph{Findings of the {Association} for
  {Computational} {Linguistics}: {EMNLP} 2022}}.
  \bibinfo{publisher}{Association for Computational Linguistics},
  \bibinfo{address}{Abu Dhabi, United Arab Emirates},
  \bibinfo{pages}{2216--2225}.
\newblock
\urldef\tempurl%
\url{https://aclanthology.org/2022.findings-emnlp.162}
\showURL{%
\tempurl}


\bibitem[\protect\citeauthoryear{Graesser, VanLehn, Rose, Jordan, and
  Harter}{Graesser et~al\mbox{.}}{2001}]%
        {graesser_intelligent_2001}
\bibfield{author}{\bibinfo{person}{Arthur~C. Graesser}, \bibinfo{person}{Kurt
  VanLehn}, \bibinfo{person}{Carolyn~P. Rose}, \bibinfo{person}{Pamela~W.
  Jordan}, {and} \bibinfo{person}{Derek Harter}.}
  \bibinfo{year}{2001}\natexlab{}.
\newblock \showarticletitle{Intelligent {Tutoring} {Systems} with
  {Conversational} {Dialogue}}.
\newblock \bibinfo{journal}{\emph{AI Magazine}} \bibinfo{volume}{22},
  \bibinfo{number}{4} (\bibinfo{date}{Dec.} \bibinfo{year}{2001}),
  \bibinfo{pages}{39--39}.
\newblock
\showISSN{2371-9621}
\urldef\tempurl%
\url{https://doi.org/10.1609/aimag.v22i4.1591}
\showDOI{\tempurl}


\bibitem[\protect\citeauthoryear{Jannin, Ganier, and De~Vries}{Jannin
  et~al\mbox{.}}{2019}]%
        {jannin_atomized_2019}
\bibfield{author}{\bibinfo{person}{Leslie Jannin}, \bibinfo{person}{Franck
  Ganier}, {and} \bibinfo{person}{Philine De~Vries}.}
  \bibinfo{year}{2019}\natexlab{}.
\newblock \showarticletitle{Atomized or {Delayed} {Execution}? {An}
  {Alternative} {Paradigm} for the {Study} of {Procedural} {Learning}}.
\newblock \bibinfo{journal}{\emph{Journal of Educational Psychology}}
  \bibinfo{volume}{111} (\bibinfo{year}{2019}), \bibinfo{pages}{1406--1415}.
\newblock
\showISSN{1939-2176}
\urldef\tempurl%
\url{https://doi.org/10.1037/edu0000357}
\showDOI{\tempurl}
\newblock
\shownote{Place: US Publisher: American Psychological Association.}


\bibitem[\protect\citeauthoryear{Kiddon, Ponnuraj, Zettlemoyer, and
  Choi}{Kiddon et~al\mbox{.}}{2015}]%
        {kiddon_mise_2015}
\bibfield{author}{\bibinfo{person}{Chloé Kiddon},
  \bibinfo{person}{Ganesa~Thandavam Ponnuraj}, \bibinfo{person}{Luke
  Zettlemoyer}, {and} \bibinfo{person}{Yejin Choi}.}
  \bibinfo{year}{2015}\natexlab{}.
\newblock \showarticletitle{Mise en {Place}: {Unsupervised} {Interpretation} of
  {Instructional} {Recipes}}. In \bibinfo{booktitle}{\emph{Proceedings of the
  2015 {Conference} on {Empirical} {Methods} in {Natural} {Language}
  {Processing}}}. \bibinfo{publisher}{Association for Computational
  Linguistics}, \bibinfo{address}{Lisbon, Portugal}, \bibinfo{pages}{982--992}.
\newblock
\urldef\tempurl%
\url{https://doi.org/10.18653/v1/D15-1114}
\showDOI{\tempurl}


\bibitem[\protect\citeauthoryear{Kim, Maddela, Kriz, Xu, and
  Callison-Burch}{Kim et~al\mbox{.}}{2021}]%
        {kim_bisect_2021}
\bibfield{author}{\bibinfo{person}{Joongwon Kim}, \bibinfo{person}{Mounica
  Maddela}, \bibinfo{person}{Reno Kriz}, \bibinfo{person}{Wei Xu}, {and}
  \bibinfo{person}{Chris Callison-Burch}.} \bibinfo{year}{2021}\natexlab{}.
\newblock \showarticletitle{{BiSECT}: {Learning} to {Split} and {Rephrase}
  {Sentences} with {Bitexts}}. In \bibinfo{booktitle}{\emph{Proceedings of the
  2021 {Conference} on {Empirical} {Methods} in {Natural} {Language}
  {Processing}}} \emph{(\bibinfo{series}{{EMNLP} '21})}.
  \bibinfo{publisher}{Association for Computational Linguistics},
  \bibinfo{address}{Online and Punta Cana, Dominican Republic},
  \bibinfo{pages}{6193--6209}.
\newblock
\urldef\tempurl%
\url{https://doi.org/10.18653/v1/2021.emnlp-main.500}
\showDOI{\tempurl}


\bibitem[\protect\citeauthoryear{Kosch, Wennrich, Topp, Muntzinger, and
  Schmidt}{Kosch et~al\mbox{.}}{2019}]%
        {kosch_digital_2019}
\bibfield{author}{\bibinfo{person}{Thomas Kosch}, \bibinfo{person}{Kevin
  Wennrich}, \bibinfo{person}{Daniel Topp}, \bibinfo{person}{Marcel
  Muntzinger}, {and} \bibinfo{person}{Albrecht Schmidt}.}
  \bibinfo{year}{2019}\natexlab{}.
\newblock \showarticletitle{The {Digital} {Cooking} {Coach}: {Using} {Visual}
  and {Auditory} {In}-{Situ} {Instructions} to {Assist} {Cognitively}
  {Impaired} during {Cooking}}. In \bibinfo{booktitle}{\emph{Proceedings of the
  12th {ACM} {International} {Conference} on {PErvasive} {Technologies}
  {Related} to {Assistive} {Environments}}} \emph{(\bibinfo{series}{{PETRA}
  '19})}. \bibinfo{publisher}{Association for Computing Machinery},
  \bibinfo{address}{New York, NY, USA}, \bibinfo{pages}{156--163}.
\newblock
\showISBNx{978-1-4503-6232-0}
\urldef\tempurl%
\url{https://doi.org/10.1145/3316782.3321524}
\showDOI{\tempurl}


\bibitem[\protect\citeauthoryear{Langevin, Lordon, Avrahami, Cowan, Hirsch, and
  Hsieh}{Langevin et~al\mbox{.}}{2021}]%
        {langevin_heuristic_2021}
\bibfield{author}{\bibinfo{person}{Raina Langevin}, \bibinfo{person}{Ross~J
  Lordon}, \bibinfo{person}{Thi Avrahami}, \bibinfo{person}{Benjamin~R. Cowan},
  \bibinfo{person}{Tad Hirsch}, {and} \bibinfo{person}{Gary Hsieh}.}
  \bibinfo{year}{2021}\natexlab{}.
\newblock \showarticletitle{Heuristic {Evaluation} of {Conversational}
  {Agents}}. In \bibinfo{booktitle}{\emph{Proceedings of the 2021 {CHI}
  {Conference} on {Human} {Factors} in {Computing} {Systems}}}
  \emph{(\bibinfo{series}{{CHI} '21})}. \bibinfo{publisher}{Association for
  Computing Machinery}, \bibinfo{address}{New York, NY, USA},
  \bibinfo{pages}{1--15}.
\newblock
\showISBNx{978-1-4503-8096-6}
\urldef\tempurl%
\url{https://doi.org/10.1145/3411764.3445312}
\showDOI{\tempurl}


\bibitem[\protect\citeauthoryear{Large, Burnett, and Clark}{Large
  et~al\mbox{.}}{2019}]%
        {large_lessons_2019}
\bibfield{author}{\bibinfo{person}{David~R. Large}, \bibinfo{person}{Gary
  Burnett}, {and} \bibinfo{person}{Leigh Clark}.}
  \bibinfo{year}{2019}\natexlab{}.
\newblock \showarticletitle{Lessons from {Oz}: {Design} {Guidelines} for
  {Automotive} {Conversational} {User} {Interfaces}}. In
  \bibinfo{booktitle}{\emph{Proceedings of the 11th {International}
  {Conference} on {Automotive} {User} {Interfaces} and {Interactive}
  {Vehicular} {Applications}: {Adjunct} {Proceedings}}}
  \emph{(\bibinfo{series}{{AutomotiveUI} '19})}.
  \bibinfo{publisher}{Association for Computing Machinery},
  \bibinfo{address}{New York, NY, USA}, \bibinfo{pages}{335--340}.
\newblock
\showISBNx{978-1-4503-6920-6}
\urldef\tempurl%
\url{https://doi.org/10.1145/3349263.3351314}
\showDOI{\tempurl}


\bibitem[\protect\citeauthoryear{Lyu, Zhang, and Callison-Burch}{Lyu
  et~al\mbox{.}}{2021}]%
        {lyu_goal-oriented_2021}
\bibfield{author}{\bibinfo{person}{Qing Lyu}, \bibinfo{person}{Li Zhang}, {and}
  \bibinfo{person}{Chris Callison-Burch}.} \bibinfo{year}{2021}\natexlab{}.
\newblock \showarticletitle{Goal-{Oriented} {Script} {Construction}}. In
  \bibinfo{booktitle}{\emph{Proceedings of the 14th {International}
  {Conference} on {Natural} {Language} {Generation}}}.
  \bibinfo{publisher}{Association for Computational Linguistics},
  \bibinfo{address}{Aberdeen, Scotland, UK}, \bibinfo{pages}{184--200}.
\newblock
\urldef\tempurl%
\url{https://aclanthology.org/2021.inlg-1.19}
\showURL{%
\tempurl}


\bibitem[\protect\citeauthoryear{Miller}{Miller}{1956}]%
        {miller_magical_1956}
\bibfield{author}{\bibinfo{person}{George~A. Miller}.}
  \bibinfo{year}{1956}\natexlab{}.
\newblock \showarticletitle{The {Magical} {Number} {Seven}, {Plus} or {Minus}
  {Two}: {Some} {Limits} on {Our} {Capacity} for {Processing} {Information}}.
\newblock \bibinfo{journal}{\emph{Psychological Review}}  \bibinfo{volume}{63}
  (\bibinfo{year}{1956}), \bibinfo{pages}{81--97}.
\newblock
\showISSN{1939-1471}
\urldef\tempurl%
\url{https://doi.org/10.1037/h0043158}
\showDOI{\tempurl}
\newblock
\shownote{Place: US Publisher: American Psychological Association.}


\bibitem[\protect\citeauthoryear{Miyashiro}{Miyashiro}{2023}]%
        {miyashiro_tuscan_2023}
\bibfield{author}{\bibinfo{person}{Lauren Miyashiro}.}
  \bibinfo{year}{2023}\natexlab{}.
\newblock \bibinfo{title}{Tuscan {Butter} {Salmon}}.
\newblock
\newblock
\urldef\tempurl%
\url{https://www.delish.com/cooking/recipe-ideas/recipes/a58412/tuscan-butter-salmon-recipe/}
\showURL{%
\tempurl}
\newblock
\shownote{Section: Recipes.}


\bibitem[\protect\citeauthoryear{Murad and Munteanu}{Murad and
  Munteanu}{2019}]%
        {murad_i_2019}
\bibfield{author}{\bibinfo{person}{Christine Murad} {and}
  \bibinfo{person}{Cosmin Munteanu}.} \bibinfo{year}{2019}\natexlab{}.
\newblock \showarticletitle{"{I} {Don}'t {Know} {What} {You}'re {Talking}
  {About}, {HALexa}": {The} {Case} for {Voice} {User} {Interface}
  {Guidelines}}. In \bibinfo{booktitle}{\emph{Proceedings of the 1st
  {International} {Conference} on {Conversational} {User} {Interfaces}}}
  \emph{(\bibinfo{series}{{CUI} '19})}. \bibinfo{publisher}{Association for
  Computing Machinery}, \bibinfo{address}{New York, NY, USA},
  \bibinfo{pages}{1--3}.
\newblock
\showISBNx{978-1-4503-7187-2}
\urldef\tempurl%
\url{https://doi.org/10.1145/3342775.3342795}
\showDOI{\tempurl}


\bibitem[\protect\citeauthoryear{Murad, Munteanu, Clark, and Cowan}{Murad
  et~al\mbox{.}}{2018}]%
        {murad_design_2018}
\bibfield{author}{\bibinfo{person}{Christine Murad}, \bibinfo{person}{Cosmin
  Munteanu}, \bibinfo{person}{Leigh Clark}, {and} \bibinfo{person}{Benjamin~R.
  Cowan}.} \bibinfo{year}{2018}\natexlab{}.
\newblock \showarticletitle{Design {Guidelines} for {Hands}-{Free} {Speech}
  {Interaction}}. In \bibinfo{booktitle}{\emph{Proceedings of the 20th
  {International} {Conference} on {Human}-{Computer} {Interaction} with
  {Mobile} {Devices} and {Services} {Adjunct}}}
  \emph{(\bibinfo{series}{{MobileHCI} '18})}. \bibinfo{publisher}{Association
  for Computing Machinery}, \bibinfo{address}{New York, NY, USA},
  \bibinfo{pages}{269--276}.
\newblock
\showISBNx{978-1-4503-5941-2}
\urldef\tempurl%
\url{https://doi.org/10.1145/3236112.3236149}
\showDOI{\tempurl}


\bibitem[\protect\citeauthoryear{Murad, Munteanu, R.~Cowan, and Clark}{Murad
  et~al\mbox{.}}{2021}]%
        {murad_finding_2021}
\bibfield{author}{\bibinfo{person}{Christine Murad}, \bibinfo{person}{Cosmin
  Munteanu}, \bibinfo{person}{Benjamin R.~Cowan}, {and} \bibinfo{person}{Leigh
  Clark}.} \bibinfo{year}{2021}\natexlab{}.
\newblock \showarticletitle{Finding a {New} {Voice}: {Transitioning}
  {Designers} from {GUI} to {VUI} {Design}}. In
  \bibinfo{booktitle}{\emph{Proceedings of the 3rd {Conference} on
  {Conversational} {User} {Interfaces}}} \emph{(\bibinfo{series}{{CUI} '21})}.
  \bibinfo{publisher}{Association for Computing Machinery},
  \bibinfo{address}{New York, NY, USA}, \bibinfo{pages}{1--12}.
\newblock
\showISBNx{978-1-4503-8998-3}
\urldef\tempurl%
\url{https://doi.org/10.1145/3469595.3469617}
\showDOI{\tempurl}


\bibitem[\protect\citeauthoryear{Norrick}{Norrick}{2011}]%
        {norrick_conversational_2011}
\bibfield{author}{\bibinfo{person}{Neal~R. Norrick}.}
  \bibinfo{year}{2011}\natexlab{}.
\newblock \showarticletitle{Conversational {Recipe} {Telling}}.
\newblock \bibinfo{journal}{\emph{Journal of Pragmatics}} \bibinfo{volume}{43},
  \bibinfo{number}{11} (\bibinfo{date}{Sept.} \bibinfo{year}{2011}),
  \bibinfo{pages}{2740--2761}.
\newblock
\showISSN{0378-2166}
\urldef\tempurl%
\url{https://doi.org/10.1016/j.pragma.2011.04.010}
\showDOI{\tempurl}


\bibitem[\protect\citeauthoryear{Nouri, Fourney, Sim, and White}{Nouri
  et~al\mbox{.}}{2019}]%
        {nouri_supporting_2019}
\bibfield{author}{\bibinfo{person}{Elnaz Nouri}, \bibinfo{person}{Adam
  Fourney}, \bibinfo{person}{Robert Sim}, {and} \bibinfo{person}{Ryen~W.
  White}.} \bibinfo{year}{2019}\natexlab{}.
\newblock \showarticletitle{Supporting {Complex} {Tasks} {Using} {Multiple}
  {Devices}}. In \bibinfo{booktitle}{\emph{Proceedings of the {Twelfth} {ACM}
  {International} {Conference} on {Web} {Search} and {Data} {Mining}}}.
  \bibinfo{publisher}{Association for Computing Machinery}.
\newblock
\urldef\tempurl%
\url{https://www.microsoft.com/en-us/research/publication/supporting-complex-tasks-using-multiple-devices/}
\showURL{%
\tempurl}


\bibitem[\protect\citeauthoryear{{OpenAI}}{{OpenAI}}{2022}]%
        {openai_chatgpt_2022}
\bibfield{author}{\bibinfo{person}{{OpenAI}}.} \bibinfo{year}{2022}\natexlab{}.
\newblock \bibinfo{title}{{ChatGPT}: {Optimizing} {Language} {Models} for
  {Dialogue}}.
\newblock
\newblock
\urldef\tempurl%
\url{https://openai.com/blog/chatgpt/}
\showURL{%
\tempurl}
\newblock
\shownote{Publication Title: OpenAI.}


\bibitem[\protect\citeauthoryear{Panagopoulou, Arora, Zhang, Cugini, You, Yang,
  Zhou, Wang, Hou, Hwang, Martin, Shi, Callison-Burch, and
  Yatskar}{Panagopoulou et~al\mbox{.}}{2022}]%
        {panagopoulou_quakerbot_2022}
\bibfield{author}{\bibinfo{person}{Artemis Panagopoulou},
  \bibinfo{person}{Manni Arora}, \bibinfo{person}{Li Zhang},
  \bibinfo{person}{Dimitri Cugini}, \bibinfo{person}{Weiqiu You},
  \bibinfo{person}{Yue Yang}, \bibinfo{person}{Liyang Zhou},
  \bibinfo{person}{Yuxuan Wang}, \bibinfo{person}{Zhaoyi Hou},
  \bibinfo{person}{Alyssa Hwang}, \bibinfo{person}{Lara Martin},
  \bibinfo{person}{Sherry Shi}, \bibinfo{person}{Chris Callison-Burch}, {and}
  \bibinfo{person}{Mark Yatskar}.} \bibinfo{year}{2022}\natexlab{}.
\newblock \showarticletitle{{QuakerBot}: {A} {Household} {Dialog} {System}
  {Powered} by {Large} {Language} {Models}}. In \bibinfo{booktitle}{\emph{Alexa
  {Prize} {TaskBot} {Challenge} {Proceedings}}}.
\newblock
\urldef\tempurl%
\url{https://www.amazon.science/alexa-prize/proceedings/quakerbot-a-household-dialog-system-powered-by-large-language-models}
\showURL{%
\tempurl}


\bibitem[\protect\citeauthoryear{Porcheron, Fischer, Reeves, and
  Sharples}{Porcheron et~al\mbox{.}}{2018}]%
        {porcheron_voice_2018}
\bibfield{author}{\bibinfo{person}{Martin Porcheron}, \bibinfo{person}{Joel~E.
  Fischer}, \bibinfo{person}{Stuart Reeves}, {and} \bibinfo{person}{Sarah
  Sharples}.} \bibinfo{year}{2018}\natexlab{}.
\newblock \showarticletitle{Voice {Interfaces} in {Everyday} {Life}}. In
  \bibinfo{booktitle}{\emph{Proceedings of the 2018 {CHI} {Conference} on
  {Human} {Factors} in {Computing} {Systems}}} \emph{(\bibinfo{series}{{CHI}
  '18})}. \bibinfo{publisher}{Association for Computing Machinery},
  \bibinfo{address}{New York, NY, USA}, \bibinfo{pages}{1--12}.
\newblock
\showISBNx{978-1-4503-5620-6}
\urldef\tempurl%
\url{https://doi.org/10.1145/3173574.3174214}
\showDOI{\tempurl}


\bibitem[\protect\citeauthoryear{Ramesh, Pavlov, Goh, Gray, Voss, Radford,
  Chen, and Sutskever}{Ramesh et~al\mbox{.}}{2021}]%
        {ramesh_zero-shot_2021}
\bibfield{author}{\bibinfo{person}{Aditya Ramesh}, \bibinfo{person}{Mikhail
  Pavlov}, \bibinfo{person}{Gabriel Goh}, \bibinfo{person}{Scott Gray},
  \bibinfo{person}{Chelsea Voss}, \bibinfo{person}{Alec Radford},
  \bibinfo{person}{Mark Chen}, {and} \bibinfo{person}{Ilya Sutskever}.}
  \bibinfo{year}{2021}\natexlab{}.
\newblock \showarticletitle{Zero-{Shot} {Text}-to-{Image} {Generation}}. In
  \bibinfo{booktitle}{\emph{Proceedings of the 38th {International}
  {Conference} on {Machine} {Learning}}} \emph{(\bibinfo{series}{Proceedings of
  {Machine} {Learning} {Research}})}, \bibfield{editor}{\bibinfo{person}{Marina
  Meila} {and} \bibinfo{person}{Tong Zhang}} (Eds.),
  Vol.~\bibinfo{volume}{139}. \bibinfo{publisher}{PMLR},
  \bibinfo{pages}{8821--8831}.
\newblock
\urldef\tempurl%
\url{https://proceedings.mlr.press/v139/ramesh21a.html}
\showURL{%
\tempurl}


\bibitem[\protect\citeauthoryear{Reif, Ippolito, Yuan, Coenen, Callison-Burch,
  and Wei}{Reif et~al\mbox{.}}{2022}]%
        {reif_recipe_2022}
\bibfield{author}{\bibinfo{person}{Emily Reif}, \bibinfo{person}{Daphne
  Ippolito}, \bibinfo{person}{Ann Yuan}, \bibinfo{person}{Andy Coenen},
  \bibinfo{person}{Chris Callison-Burch}, {and} \bibinfo{person}{Jason Wei}.}
  \bibinfo{year}{2022}\natexlab{}.
\newblock \showarticletitle{A {Recipe} for {Arbitrary} {Text} {Style}
  {Transfer} with {Large} {Language} {Models}}. In
  \bibinfo{booktitle}{\emph{Proceedings of the 60th {Annual} {Meeting} of the
  {Association} for {Computational} {Linguistics} ({Volume} 2: {Short}
  {Papers})}} \emph{(\bibinfo{series}{{ACL} '22})}.
  \bibinfo{publisher}{Association for Computational Linguistics},
  \bibinfo{address}{Dublin, Ireland}, \bibinfo{pages}{837--848}.
\newblock
\urldef\tempurl%
\url{https://doi.org/10.18653/v1/2022.acl-short.94}
\showDOI{\tempurl}


\bibitem[\protect\citeauthoryear{Rombach, Blattmann, Lorenz, Esser, and
  Ommer}{Rombach et~al\mbox{.}}{2022}]%
        {rombach_high-resolution_2022}
\bibfield{author}{\bibinfo{person}{Robin Rombach}, \bibinfo{person}{Andreas
  Blattmann}, \bibinfo{person}{Dominik Lorenz}, \bibinfo{person}{Patrick
  Esser}, {and} \bibinfo{person}{Björn Ommer}.}
  \bibinfo{year}{2022}\natexlab{}.
\newblock \showarticletitle{High-{Resolution} {Image} {Synthesis} {With}
  {Latent} {Diffusion} {Models}} \emph{(\bibinfo{series}{{CVPR} '22})}.
  \bibinfo{pages}{10684--10695}.
\newblock
\urldef\tempurl%
\url{https://openaccess.thecvf.com/content/CVPR2022/html/Rombach_High-Resolution_Image_Synthesis_With_Latent_Diffusion_Models_CVPR_2022_paper.html}
\showURL{%
\tempurl}


\bibitem[\protect\citeauthoryear{Sato, Watanabe, and Rekimoto}{Sato
  et~al\mbox{.}}{2014}]%
        {sato_mimicook_2014}
\bibfield{author}{\bibinfo{person}{Ayaka Sato}, \bibinfo{person}{Keita
  Watanabe}, {and} \bibinfo{person}{Jun Rekimoto}.}
  \bibinfo{year}{2014}\natexlab{}.
\newblock \showarticletitle{{MimiCook}: {A} {Cooking} {Assistant} {System} with
  {Situated} {Guidance}}. In \bibinfo{booktitle}{\emph{Proceedings of the 8th
  {International} {Conference} on {Tangible}, {Embedded} and {Embodied}
  {Interaction}}} \emph{(\bibinfo{series}{{TEI} '14})}.
  \bibinfo{publisher}{Association for Computing Machinery},
  \bibinfo{address}{New York, NY, USA}, \bibinfo{pages}{121--124}.
\newblock
\showISBNx{978-1-4503-2635-3}
\urldef\tempurl%
\url{https://doi.org/10.1145/2540930.2540952}
\showDOI{\tempurl}


\bibitem[\protect\citeauthoryear{Schoop, Nguyen, Lim, Savage, Follmer, and
  Hartmann}{Schoop et~al\mbox{.}}{2016}]%
        {schoop_drill_2016}
\bibfield{author}{\bibinfo{person}{Eldon Schoop}, \bibinfo{person}{Michelle
  Nguyen}, \bibinfo{person}{Daniel Lim}, \bibinfo{person}{Valkyrie Savage},
  \bibinfo{person}{Sean Follmer}, {and} \bibinfo{person}{Björn Hartmann}.}
  \bibinfo{year}{2016}\natexlab{}.
\newblock \showarticletitle{Drill {Sergeant}: {Supporting} {Physical}
  {Construction} {Projects} through an {Ecosystem} of {Augmented} {Tools}}. In
  \bibinfo{booktitle}{\emph{Proceedings of the 2016 {CHI} {Conference}
  {Extended} {Abstracts} on {Human} {Factors} in {Computing} {Systems}}}
  \emph{(\bibinfo{series}{{CHI} {EA} '16})}. \bibinfo{publisher}{Association
  for Computing Machinery}, \bibinfo{address}{New York, NY, USA},
  \bibinfo{pages}{1607--1614}.
\newblock
\showISBNx{978-1-4503-4082-3}
\urldef\tempurl%
\url{https://doi.org/10.1145/2851581.2892429}
\showDOI{\tempurl}


\bibitem[\protect\citeauthoryear{Sciuto, Saini, Forlizzi, and Hong}{Sciuto
  et~al\mbox{.}}{2018}]%
        {sciuto_hey_2018}
\bibfield{author}{\bibinfo{person}{Alex Sciuto}, \bibinfo{person}{Arnita
  Saini}, \bibinfo{person}{Jodi Forlizzi}, {and} \bibinfo{person}{Jason~I.
  Hong}.} \bibinfo{year}{2018}\natexlab{}.
\newblock \showarticletitle{"{Hey} {Alexa}, {What}'s {Up}?": {A}
  {Mixed}-{Methods} {Studies} of {In}-{Home} {Conversational} {Agent} {Usage}}.
  In \bibinfo{booktitle}{\emph{Proceedings of the 2018 {Designing}
  {Interactive} {Systems} {Conference}}} \emph{(\bibinfo{series}{{DIS} '18})}.
  \bibinfo{publisher}{Association for Computing Machinery},
  \bibinfo{address}{New York, NY, USA}, \bibinfo{pages}{857--868}.
\newblock
\showISBNx{978-1-4503-5198-0}
\urldef\tempurl%
\url{https://doi.org/10.1145/3196709.3196772}
\showDOI{\tempurl}


\bibitem[\protect\citeauthoryear{Sherwani, Yu, Paek, Czerwinski, Ju, and
  Acero}{Sherwani et~al\mbox{.}}{2007}]%
        {sherwani_voicepedia_2007}
\bibfield{author}{\bibinfo{person}{J. Sherwani}, \bibinfo{person}{Dong Yu},
  \bibinfo{person}{Tim Paek}, \bibinfo{person}{M. Czerwinski},
  \bibinfo{person}{Y. Ju}, {and} \bibinfo{person}{A. Acero}.}
  \bibinfo{year}{2007}\natexlab{}.
\newblock \showarticletitle{Voicepedia: {Towards} {Speech}-{Based} {Access} to
  {Unstructured} {Information}}. \bibinfo{publisher}{ISCA},
  \bibinfo{pages}{146--149}.
\newblock
\urldef\tempurl%
\url{https://www.semanticscholar.org/paper/Voicepedia%3A-towards-speech-based-access-to-Sherwani-Yu/dd6f19b38072b2dad6633bd31879b7f9e7138dcd}
\showURL{%
\tempurl}


\bibitem[\protect\citeauthoryear{Speer, Chin, and Havasi}{Speer
  et~al\mbox{.}}{2017}]%
        {speer_conceptnet_2017}
\bibfield{author}{\bibinfo{person}{Robyn Speer}, \bibinfo{person}{Joshua Chin},
  {and} \bibinfo{person}{Catherine Havasi}.} \bibinfo{year}{2017}\natexlab{}.
\newblock \showarticletitle{{ConceptNet} 5.5: {An} {Open} {Multilingual}
  {Graph} of {General} {Knowledge}}. In \bibinfo{booktitle}{\emph{Proceedings
  of the {Thirty}-{First} {AAAI} {Conference} on {Artificial} {Intelligence}}}
  \emph{(\bibinfo{series}{{AAAI} '17})}. \bibinfo{publisher}{AAAI Press},
  \bibinfo{address}{San Francisco, California, USA},
  \bibinfo{pages}{4444--4451}.
\newblock


\bibitem[\protect\citeauthoryear{Tabbers, Martens, and van
  Merriënboer}{Tabbers et~al\mbox{.}}{2004}]%
        {tabbers_multimedia_2004}
\bibfield{author}{\bibinfo{person}{Huib~K. Tabbers}, \bibinfo{person}{Rob~L.
  Martens}, {and} \bibinfo{person}{Jeroen J.~G. van Merriënboer}.}
  \bibinfo{year}{2004}\natexlab{}.
\newblock \showarticletitle{Multimedia {Instructions} and {Cognitive} {Load}
  {Theory}: {Effects} of {Modality} and {Cueing}}.
\newblock \bibinfo{journal}{\emph{The British Journal of Educational
  Psychology}} \bibinfo{volume}{74}, \bibinfo{number}{Pt 1}
  (\bibinfo{date}{March} \bibinfo{year}{2004}), \bibinfo{pages}{71--81}.
\newblock
\showISSN{0007-0998}
\urldef\tempurl%
\url{https://doi.org/10.1348/000709904322848824}
\showDOI{\tempurl}


\bibitem[\protect\citeauthoryear{Tu, Rim, and Pustejovsky}{Tu
  et~al\mbox{.}}{2022}]%
        {tu_competence-based_2022}
\bibfield{author}{\bibinfo{person}{Jingxuan Tu}, \bibinfo{person}{Kyeongmin
  Rim}, {and} \bibinfo{person}{James Pustejovsky}.}
  \bibinfo{year}{2022}\natexlab{}.
\newblock \showarticletitle{Competence-based {Question} {Generation}}. In
  \bibinfo{booktitle}{\emph{Proceedings of the 29th {International}
  {Conference} on {Computational} {Linguistics}}}
  \emph{(\bibinfo{series}{{COLING} '22})}. \bibinfo{publisher}{International
  Committee on Computational Linguistics}, \bibinfo{address}{Gyeongju, Republic
  of Korea}, \bibinfo{pages}{1521--1533}.
\newblock
\urldef\tempurl%
\url{https://aclanthology.org/2022.coling-1.131}
\showURL{%
\tempurl}


\bibitem[\protect\citeauthoryear{Vtyurina and Fourney}{Vtyurina and
  Fourney}{2018}]%
        {vtyurina_exploring_2018}
\bibfield{author}{\bibinfo{person}{Alexandra Vtyurina} {and}
  \bibinfo{person}{Adam Fourney}.} \bibinfo{year}{2018}\natexlab{}.
\newblock \showarticletitle{Exploring the {Role} of {Conversational} {Cues} in
  {Guided} {Task} {Support} with {Virtual} {Assistants}}. In
  \bibinfo{booktitle}{\emph{Proceedings of the 2018 {CHI} {Conference} on
  {Human} {Factors} in {Computing} {Systems}}} \emph{(\bibinfo{series}{{CHI}
  '18})}. \bibinfo{publisher}{Association for Computing Machinery},
  \bibinfo{address}{New York, NY, USA}, \bibinfo{pages}{1--7}.
\newblock
\showISBNx{978-1-4503-5620-6}
\urldef\tempurl%
\url{https://doi.org/10.1145/3173574.3173782}
\showDOI{\tempurl}


\bibitem[\protect\citeauthoryear{Völkel, Buschek, Eiband, Cowan, and
  Hussmann}{Völkel et~al\mbox{.}}{2021}]%
        {volkel_eliciting_2021}
\bibfield{author}{\bibinfo{person}{Sarah~Theres Völkel},
  \bibinfo{person}{Daniel Buschek}, \bibinfo{person}{Malin Eiband},
  \bibinfo{person}{Benjamin~R. Cowan}, {and} \bibinfo{person}{Heinrich
  Hussmann}.} \bibinfo{year}{2021}\natexlab{}.
\newblock \showarticletitle{Eliciting and {Analysing} {Users}’ {Envisioned}
  {Dialogues} with {Perfect} {Voice} {Assistants}}. In
  \bibinfo{booktitle}{\emph{Proceedings of the 2021 {CHI} {Conference} on
  {Human} {Factors} in {Computing} {Systems}}} \emph{(\bibinfo{series}{{CHI}
  '21})}. \bibinfo{publisher}{Association for Computing Machinery},
  \bibinfo{address}{New York, NY, USA}, \bibinfo{pages}{1--15}.
\newblock
\showISBNx{978-1-4503-8096-6}
\urldef\tempurl%
\url{https://doi.org/10.1145/3411764.3445536}
\showDOI{\tempurl}


\bibitem[\protect\citeauthoryear{Winkler, Söllner, Neuweiler, Conti~Rossini,
  and Leimeister}{Winkler et~al\mbox{.}}{2019}]%
        {winkler_alexa_2019}
\bibfield{author}{\bibinfo{person}{Rainer Winkler}, \bibinfo{person}{Matthias
  Söllner}, \bibinfo{person}{Maya~Lisa Neuweiler}, \bibinfo{person}{Flavia
  Conti~Rossini}, {and} \bibinfo{person}{Jan~Marco Leimeister}.}
  \bibinfo{year}{2019}\natexlab{}.
\newblock \showarticletitle{Alexa, {Can} {You} {Help} {Us} {Solve} {This}
  {Problem}? {How} {Conversations} {With} {Smart} {Personal} {Assistant}
  {Tutors} {Increase} {Task} {Group} {Outcomes}}. In
  \bibinfo{booktitle}{\emph{Extended {Abstracts} of the 2019 {CHI} {Conference}
  on {Human} {Factors} in {Computing} {Systems}}} \emph{(\bibinfo{series}{{CHI}
  {EA} '19})}. \bibinfo{publisher}{Association for Computing Machinery},
  \bibinfo{address}{New York, NY, USA}, \bibinfo{pages}{1--6}.
\newblock
\showISBNx{978-1-4503-5971-9}
\urldef\tempurl%
\url{https://doi.org/10.1145/3290607.3313090}
\showDOI{\tempurl}


\bibitem[\protect\citeauthoryear{Wu, Hoi, Socher, and Xiong}{Wu
  et~al\mbox{.}}{2020}]%
        {wu_tod-bert_2020}
\bibfield{author}{\bibinfo{person}{Chien-Sheng Wu},
  \bibinfo{person}{Steven~C.H. Hoi}, \bibinfo{person}{Richard Socher}, {and}
  \bibinfo{person}{Caiming Xiong}.} \bibinfo{year}{2020}\natexlab{}.
\newblock \showarticletitle{{TOD}-{BERT}: {Pre}-trained {Natural} {Language}
  {Understanding} for {Task}-{Oriented} {Dialogue}}. In
  \bibinfo{booktitle}{\emph{Proceedings of the 2020 {Conference} on {Empirical}
  {Methods} in {Natural} {Language} {Processing} ({EMNLP})}}.
  \bibinfo{publisher}{Association for Computational Linguistics},
  \bibinfo{address}{Online}, \bibinfo{pages}{917--929}.
\newblock
\urldef\tempurl%
\url{https://doi.org/10.18653/v1/2020.emnlp-main.66}
\showDOI{\tempurl}


\bibitem[\protect\citeauthoryear{Yang, Kim, Panagopoulou, Yatskar, and
  Callison-Burch}{Yang et~al\mbox{.}}{2021a}]%
        {yang_induce_2021}
\bibfield{author}{\bibinfo{person}{Yue Yang}, \bibinfo{person}{Joongwon Kim},
  \bibinfo{person}{Artemis Panagopoulou}, \bibinfo{person}{Mark Yatskar}, {and}
  \bibinfo{person}{Chris Callison-Burch}.} \bibinfo{year}{2021}\natexlab{a}.
\newblock \showarticletitle{Induce, {Edit}, {Retrieve}: {Language} {Grounded}
  {Multimodal} {Schema} for {Instructional} {Video} {Retrieval}}.
\newblock \bibinfo{journal}{\emph{Computing Research Repository}}
  \bibinfo{volume}{abs/2111.09276} (\bibinfo{year}{2021}).
\newblock
\urldef\tempurl%
\url{https://arxiv.org/abs/2111.09276}
\showURL{%
\tempurl}
\newblock
\shownote{arXiv: 2111.09276.}


\bibitem[\protect\citeauthoryear{Yang, Panagopoulou, Lyu, Zhang, Yatskar, and
  Callison-Burch}{Yang et~al\mbox{.}}{2021b}]%
        {yang_visual_2021}
\bibfield{author}{\bibinfo{person}{Yue Yang}, \bibinfo{person}{Artemis
  Panagopoulou}, \bibinfo{person}{Qing Lyu}, \bibinfo{person}{Li Zhang},
  \bibinfo{person}{Mark Yatskar}, {and} \bibinfo{person}{Chris
  Callison-Burch}.} \bibinfo{year}{2021}\natexlab{b}.
\newblock \showarticletitle{Visual {Goal}-{Step} {Inference} using {WikiHow}}.
  In \bibinfo{booktitle}{\emph{Proceedings of the 2021 {Conference} on
  {Empirical} {Methods} in {Natural} {Language} {Processing}}}
  \emph{(\bibinfo{series}{{EMNLP} '21})}. \bibinfo{publisher}{Association for
  Computational Linguistics}, \bibinfo{address}{Online and Punta Cana,
  Dominican Republic}, \bibinfo{pages}{2167--2179}.
\newblock
\urldef\tempurl%
\url{https://doi.org/10.18653/v1/2021.emnlp-main.165}
\showDOI{\tempurl}


\bibitem[\protect\citeauthoryear{Yankelovich and Baatz}{Yankelovich and
  Baatz}{1994}]%
        {yankelovich_speechacts_1994}
\bibfield{author}{\bibinfo{person}{Nicole Yankelovich} {and}
  \bibinfo{person}{Eric Baatz}.} \bibinfo{year}{1994}\natexlab{}.
\newblock \showarticletitle{{SpeechActs}: {A} {Framework} for {Building}
  {Speech} {Applications}}. In \bibinfo{booktitle}{\emph{{AVIOS} '94
  {Conference} {Proceedings}}}. \bibinfo{publisher}{Sun Microsystems
  Laboratories, Inc.}, \bibinfo{pages}{20--23}.
\newblock
\urldef\tempurl%
\url{https://citeseerx.ist.psu.edu/document?repid=rep1&type=pdf&doi=e785ab9af208afcafe9ef1a222495df745033155}
\showURL{%
\tempurl}


\bibitem[\protect\citeauthoryear{Yankelovich, Levow, and Marx}{Yankelovich
  et~al\mbox{.}}{1995}]%
        {yankelovich_designing_1995}
\bibfield{author}{\bibinfo{person}{Nicole Yankelovich},
  \bibinfo{person}{Gina-Anne Levow}, {and} \bibinfo{person}{Matt Marx}.}
  \bibinfo{year}{1995}\natexlab{}.
\newblock \showarticletitle{Designing {SpeechActs}: {Issues} in {Speech} {User}
  {Interfaces}}. In \bibinfo{booktitle}{\emph{Proceedings of the 1995 {CHI}
  {Conference} on {Human} {Factors} in {Computing} {Systems}}}
  \emph{(\bibinfo{series}{{CHI} '95})}. \bibinfo{pages}{369--376}.
\newblock


\bibitem[\protect\citeauthoryear{Zhang, Lyu, and Callison-Burch}{Zhang
  et~al\mbox{.}}{2020a}]%
        {zhang_reasoning_2020}
\bibfield{author}{\bibinfo{person}{Li Zhang}, \bibinfo{person}{Qing Lyu}, {and}
  \bibinfo{person}{Chris Callison-Burch}.} \bibinfo{year}{2020}\natexlab{a}.
\newblock \showarticletitle{Reasoning about {Goals}, {Steps}, and {Temporal}
  {Ordering} with {WikiHow}}. In \bibinfo{booktitle}{\emph{Proceedings of the
  2020 {Conference} on {Empirical} {Methods} in {Natural} {Language}
  {Processing} ({EMNLP})}} \emph{(\bibinfo{series}{{EMNLP} '20})}.
  \bibinfo{publisher}{Association for Computational Linguistics},
  \bibinfo{address}{Online}, \bibinfo{pages}{4630--4639}.
\newblock
\urldef\tempurl%
\url{https://doi.org/10.18653/v1/2020.emnlp-main.374}
\showDOI{\tempurl}


\bibitem[\protect\citeauthoryear{Zhang, Zhu, Brahma, and Li}{Zhang
  et~al\mbox{.}}{2020b}]%
        {zhang_small_2020}
\bibfield{author}{\bibinfo{person}{Li Zhang}, \bibinfo{person}{Huaiyu Zhu},
  \bibinfo{person}{Siddhartha Brahma}, {and} \bibinfo{person}{Yunyao Li}.}
  \bibinfo{year}{2020}\natexlab{b}.
\newblock \showarticletitle{Small but {Mighty}: {New} {Benchmarks} for {Split}
  and {Rephrase}}. In \bibinfo{booktitle}{\emph{Proceedings of the 2020
  {Conference} on {Empirical} {Methods} in {Natural} {Language} {Processing}
  ({EMNLP})}} \emph{(\bibinfo{series}{{EMNLP} '20})}.
  \bibinfo{publisher}{Association for Computational Linguistics},
  \bibinfo{address}{Online}, \bibinfo{pages}{1198--1205}.
\newblock
\urldef\tempurl%
\url{https://doi.org/10.18653/v1/2020.emnlp-main.91}
\showDOI{\tempurl}


\bibitem[\protect\citeauthoryear{Zhang, Zhang, Li, and Smola}{Zhang
  et~al\mbox{.}}{2022}]%
        {zhang_automatic_2022}
\bibfield{author}{\bibinfo{person}{Zhuosheng Zhang}, \bibinfo{person}{Aston
  Zhang}, \bibinfo{person}{Mu Li}, {and} \bibinfo{person}{Alex Smola}.}
  \bibinfo{year}{2022}\natexlab{}.
\newblock \bibinfo{title}{Automatic {Chain} of {Thought} {Prompting} in {Large}
  {Language} {Models}}.
\newblock
\newblock
\urldef\tempurl%
\url{https://doi.org/10.48550/arXiv.2210.03493}
\showDOI{\tempurl}
\newblock
\shownote{arXiv:2210.03493 [cs].}


\bibitem[\protect\citeauthoryear{Zhao, Jaber, McMillan, and Munteanu}{Zhao
  et~al\mbox{.}}{2022}]%
        {zhao_rewind_2022}
\bibfield{author}{\bibinfo{person}{Yaxi Zhao}, \bibinfo{person}{Razan Jaber},
  \bibinfo{person}{Donald McMillan}, {and} \bibinfo{person}{Cosmin Munteanu}.}
  \bibinfo{year}{2022}\natexlab{}.
\newblock \showarticletitle{“{Rewind} to the {Jiggling} {Meat} {Part}”:
  {Understanding} {Voice} {Control} of {Instructional} {Videos} in {Everyday}
  {Tasks}}. In \bibinfo{booktitle}{\emph{Proceedings of the 2022 {CHI}
  {Conference} on {Human} {Factors} in {Computing} {Systems}}}
  \emph{(\bibinfo{series}{{CHI} '22})}. \bibinfo{publisher}{Association for
  Computing Machinery}, \bibinfo{address}{New York, NY, USA},
  \bibinfo{pages}{1--11}.
\newblock
\showISBNx{978-1-4503-9157-3}
\urldef\tempurl%
\url{https://doi.org/10.1145/3491102.3502036}
\showDOI{\tempurl}


\bibitem[\protect\citeauthoryear{Zhong, Liu, Ge, Mao, Jiao, Zhang, Xu, Zhu,
  Zeng, and Han}{Zhong et~al\mbox{.}}{2022}]%
        {zhong_unsupervised_2022}
\bibfield{author}{\bibinfo{person}{Ming Zhong}, \bibinfo{person}{Yang Liu},
  \bibinfo{person}{Suyu Ge}, \bibinfo{person}{Yuning Mao},
  \bibinfo{person}{Yizhu Jiao}, \bibinfo{person}{Xingxing Zhang},
  \bibinfo{person}{Yichong Xu}, \bibinfo{person}{Chenguang Zhu},
  \bibinfo{person}{Michael Zeng}, {and} \bibinfo{person}{Jiawei Han}.}
  \bibinfo{year}{2022}\natexlab{}.
\newblock \showarticletitle{Unsupervised {Multi}-{Granularity}
  {Summarization}}. In \bibinfo{booktitle}{\emph{Findings of the {Association}
  for {Computational} {Linguistics}: {EMNLP} 2022}}
  \emph{(\bibinfo{series}{Findings '22})}. \bibinfo{publisher}{Association for
  Computational Linguistics}, \bibinfo{address}{Abu Dhabi, United Arab
  Emirates}, \bibinfo{pages}{4980--4995}.
\newblock
\urldef\tempurl%
\url{https://aclanthology.org/2022.findings-emnlp.366}
\showURL{%
\tempurl}


\bibitem[\protect\citeauthoryear{Zhou, Zhang, Yang, Lyu, Yin, Callison-Burch,
  and Neubig}{Zhou et~al\mbox{.}}{2022}]%
        {zhou_show_2022}
\bibfield{author}{\bibinfo{person}{Shuyan Zhou}, \bibinfo{person}{Li Zhang},
  \bibinfo{person}{Yue Yang}, \bibinfo{person}{Qing Lyu},
  \bibinfo{person}{Pengcheng Yin}, \bibinfo{person}{Chris Callison-Burch},
  {and} \bibinfo{person}{Graham Neubig}.} \bibinfo{year}{2022}\natexlab{}.
\newblock \showarticletitle{Show {Me} {More} {Details}: {Discovering}
  {Hierarchies} of {Procedures} from {Semi}-structured {Web} {Data}}. In
  \bibinfo{booktitle}{\emph{Proceedings of the 60th {Annual} {Meeting} of the
  {Association} for {Computational} {Linguistics} ({Volume} 1: {Long}
  {Papers})}} \emph{(\bibinfo{series}{{ACL} '22})}.
  \bibinfo{publisher}{Association for Computational Linguistics},
  \bibinfo{address}{Dublin, Ireland}, \bibinfo{pages}{2998--3012}.
\newblock
\urldef\tempurl%
\url{https://doi.org/10.18653/v1/2022.acl-long.214}
\showDOI{\tempurl}


\end{thebibliography}

\appendix
\section{Inter-Rater Reliability}

\label{app:validation}
\begin{table}[H]
\begin{tabular}{@{}p{5.8cm}cc@{}}
\toprule
Event                                       & $\alpha$ & \# Excerpts \\ \midrule
Participant addresses Alexa                   & 0.99                 & 69                   \\
Participant uses wake word                 & 0.98                 & 53                   \\
Alexa responded to participant             & 0.98                 & 66                   \\
Participant selects recipe                             & 1.0                  & 2                    \\
Participant requests list of ingredients                   & 1.0                  & 5                    \\
Participant requests Alexa to start instructions                 & 1.0                  & 3                    \\
Participant requests Alexa to use a timer                    & 0.92                 & 21                   \\
Participant interrupts Alexa               & 1.0                  & 4                    \\
Participant requests the next step                                       & 0.98                 & 28                   \\
Participant requests the previous step                                   & 1.0                  & 2                    \\
Participant requests Alexa to jump to a step                      & N/A                  & 0                    \\
Participant requests Alexa to repeat                                     & 0.80                 & 3                    \\
Participant asks Alexa to stop talking & 1.0                  & 4                    \\
Participant asks Alexa for more information       & 1.0                  & 5                    \\ \bottomrule
\end{tabular}
\Description[Inter-rater reliability]{Krippendorff's alpha and number of excerpts for each event in a transcript.}
\caption{Inter-rater reliability for coding of conversational turns in Section~\ref{sec:analysis}. \textmd{For each event, the table shows the name of the code (``Event''), agreement measured by Krippendorff's alpha ($\alpha$) between two authors, and number of excerpts in the validation set of two transcripts that received that code (``\# Excerpts'').}}
\label{tab:validation}
\end{table}

\end{document}